\documentclass[pre,twocolumn,nofootinbib,twoside,showpacs,superscriptaddress,tightenlines]{revtex4-1}
\usepackage{graphicx,amssymb,amsmath,epsf,bm}
\usepackage{tabulary}
\usepackage{tabularx}
\usepackage{cancel}
\epsfclipon
\usepackage[normalem]{ulem}
\usepackage{color}
\definecolor{nred}{RGB}{224,0,0}
\definecolor{nblue}  {RGB}{28,130,185}
\definecolor{dgreen} {RGB}{78,138,21}
\definecolor{norange}{RGB}{230,120,20}

\newcommand{\be}{\begin{equation}}
\newcommand{\ee}{\end{equation}}

\begin{document}

\title{Precision calculation of universal amplitude ratios in $O(N)$ universality classes: Derivative Expansion results at order $\mathcal{O}(\partial^4)$}

\author{Gonzalo De Polsi}
\email{gdepolsi@fisica.edu.uy}
\affiliation{Instituto de F\'isica, Facultad de Ciencias, Universidad de la
	Rep\'ublica, Igu\'a 4225, 11400, Montevideo, Uruguay}
\affiliation{Instituto de F\'isica, Facultad de Ingenier\'ia, Universidad de la
	Rep\'ublica, J.H.y Reissig 565, 11300 Montevideo, Uruguay}

\author{Guzm\'an Hern\'andez-Chifflet}
\affiliation{Instituto de F\'isica, Facultad de Ingenier\'ia, Universidad de la
	Rep\'ublica, J.H.y Reissig 565, 11300 Montevideo, Uruguay}

\author{Nicol\'as Wschebor}
\affiliation{Instituto de F\'isica, Facultad de Ingenier\'ia, Universidad de la
	Rep\'ublica, J.H.y Reissig 565, 11300 Montevideo, Uruguay}

\begin{abstract}
In the last few years the derivative expansion of the Non-Perturbative Renormalization Group has proven to be a very efficient tool for the precise computation of critical quantities. In particular, recent progress in the understanding of its convergence properties allowed for an estimate of the error bars as well as the precise computation
of many critical quantities. In this work we extend previous studies to the computation of several universal amplitude ratios for the critical regime of $O(N)$ models using the derivative expansion of the Non-Perturbative Renormalization Group at order $\mathcal{O}(\partial^4)$ for three dimensional systems.
\end{abstract}

\maketitle
\section{Introduction}

The theoretical analysis of critical phenomena has made significant progress in recent  years. This has been associated with advances in at least four different theoretical approaches. Among those progresses, let us first mention the important progress generated by the development of the three-dimensional Conformal Bootstrap \cite{ElShowk:2012ht,El-Showk:2014dwa,Kos:2014bka}. This made possible to calculate exponents in the three-dimensional Ising universality class with unprecedented precision.\footnote{Recently the Conformal Bootstrap also managed to achieve a precision comparable to Monte-Carlo simulations in the $O(2)$ universality class \cite{Chester:2019ifh} with a similar numerical cost.} More recently, significant progress took place in three other types of theoretical tools. On the perturbative side, seventh order calculations have been achieved within the $\epsilon-$ expansion techniques that improve previous sixth order results \cite{Schnetz:2016fhy,Kompaniets:2017yct,Shalaby:2020faz,Abhignan:2020xcj,Shalaby:2020xvv}. In parallel, a significant improvement in the precision has been achieved within Monte-Carlo techniques \cite{Hasenbusch:2019jkj,Hasenbusch:2020pwj,Hasenbusch:2021tei}. Finally, and this will be central for the present work, a qualitative and quantitative change in the quality of the results coming from the Derivative Expansion (DE) of the Non-Perturbative Renormalization Group (NPRG) have been achieved recently \cite{Balog:2019rrg,DePolsi:2020pjk}.

Let us discuss this last point in some detail. The NPRG can be seen as a modern version of Wilson's renormalization group \cite{Wilson:1973jj,Polchinski:1983gv}. At odds with classical exact Wilsonian renormalization group equations, the NPRG determines the flow of the Gibbs free energy, which only includes one particle irreducible contributions \cite{Dupuis:2020fhh}. This technique has a long history of successes but these remained controversial because the approximations implemented within this framework were seen as uncontrolled. In relation to this, proper estimates of error bars of the results were usually not discussed. Recently, this situation has drastically changed with strong evidence supporting that the most employed approximation scheme in this context (the DE) has a finite radius of convergence \cite{Balog:2019rrg,DePolsi:2020pjk}. This is associated with an almost model-independent ``small parameter'' that varies in the range $1/9$--$1/4$ and that characterizes the improvement of precision in critical properties at successive orders of the DE \cite{Balog:2019rrg}. This allowed for the estimate of controlled error bars \cite{DePolsi:2020pjk}. The resulting precision for critical exponents turned out to be competitive with the most precise methods in the literature at third order of the DE (usually denoted $\mathcal{O}(\partial^4)$). From a practical point of view, the DE has been implemented for the calculation of critical exponents in the $O(N)$ universality classes at order $\mathcal{O}(\partial^4)$ achieving very precise and accurate results (reaching the world-best determination of critical exponents in some cases) \cite{DePolsi:2020pjk,Peli:2020yiz}. In the Ising universality class this analysis has been pushed to order $\mathcal{O}(\partial^6)$ with even higher precision \cite{Balog:2019rrg}, of course, not as precise as conformal invariance techniques \cite{El-Showk:2014dwa} but better than all other field-theoretical techniques.

The aforementioned progresses have been centered on the calculation of universal properties dominated by critical points. They are characterized by the presence of scale and (in most cases) conformal invariance.\footnote{For three-dimensional Ising and $O(N)$ models considered in the present work,  proofs of the presence of conformal symmetry at the critical point can be found in Refs.~\cite{delamotte2016scale,DePolsi2019}. An extremely strong indication of the presence of conformal symmetry in the Ising universality class is the success of the Conformal Bootstrap for this universality class \cite{ElShowk:2012ht,El-Showk:2014dwa,Kos:2014bka}.}. 
Typical examples of such quantities are critical exponents. However, there are many universal properties that are associated with the neighbourhood of the critical point (the critical domain) which are not, strictly speaking, determined by the critical point alone. In the language of the renormalization group, there are universal properties that are not determined by the fixed point. The most important and well-studied examples of such quantities are the universal amplitude ratios (UAR). Many UARs are intimately related to the universal equation of state. They are not dominated by the critical fixed point alone but include also information from the full renormalization group trajectory
going from the fixed point to the high (or low) temperature region in the coupling constant space, more pecisely joining the Wilson-Fisher fixed point to the high or low temperature fixed point. As a consequence, at least in principle, they cannot be determined from techniques that are based on conformal symmetry alone.  Indeed, methods based exclusively on conformal symmetry are extremely powerful to tackle the strictly critical properties but are not able to handle the behaviour of the theory away from the critical point. This is different from perturbation theory, Monte-Carlo methods or the NPRG where, at least in principle, all properties of the theory can be calculated (as long as the employed approximations remain valid). The simplest example is the calculation of non-universal properties such as critical temperatures or the full phase diagram. Of course, in order to compare with actual experiments, this requires the use of a realistic microscopic Hamiltonian. We consider in the present article UARs that are in an intermediate situation: they are universal, so one does not need a precise knowledge of the microscopic Hamiltonian, but they they are sensitive to the physics of the systems under study beyond the critical point.

In the present article we perform the calculation of many UARs in $O(N)$ universality classes employing the DE of the NPRG at order  $\mathcal{O}(\partial^4)$.
We employ essentially the same techniques and numerical codes as in Ref.~\cite{DePolsi:2020pjk}. The main difference is that we need now to follow the renormalization group flow along a trajectory going very close to the Wilson-Fisher fixed point and departing to the high temperature fixed point. In principle, one should also study the trajectories that depart towards the low temperature fixed point. However,  it is much more difficult to numerically control these trajectories since the corresponding free energy eventually  becomes nonanalytic because of the presence of the coexistence curve. The direct numerical approach to the low temperature phase has been successfully studied within the NPRG \cite{Berges:2000ew,Dupuis:2020fhh,Pelaez:2015nsa} at first and second order of the DE. However, in the present study, instead of studying the trajectories going from the critical fixed point towards the low temperature one, we employ a standard technique in the literature (here denoted ``parametrization technique'') in order to extract low temperature amplitudes from high temperature ones. This simplifies tremendously the numerical study but, as will be discussed below, turns out to be the principal source of error\footnote{As discussed below, uncertainties associated to the parametrization technique are, generically speaking, of the same order of magnitude than those of DE at order $\mathcal{O}(\partial^2)$.} in our calculation of amplitudes related to the low temperature phase at order $\mathcal{O}(\partial^4)$. It is clear that avoiding the use of this technique becomes a natural extension of the present work but the associated numerical cost goes beyond the present study and 
it will be considered in a forthcoming work. 

The existence of a small parameter for the DE is employed here in order to implement controlled error bars as proposed in \cite{DePolsi:2020pjk}. The obtained results have the level of precision and accuracy of the best estimates in the literature for all UARs studied  and, in some cases, achieving an even higher level of precision. 

Before finishing the introduction, let us mention that UARs have been studied with the NPRG in the pioneering works of Refs.~\cite{Tetradis94,Berges:1995mw,Berges:1996ja,Tetradis:1997bz,Berges:2000ew,PhysRevE.88.012113}. In particular, the critical equation of state has been studied both for second and for weakly first order phase transitions by using the DE of the NPRG. Let us point out, however, that these results were obtained in the early period of development of the NPRG. They have been done at leading $\mathcal{O}(\partial^0)$ or LPA order (or, in some cases, in the LPA' approximation that includes the anomalous dimension of the field) and do not include the now standard $\mathcal{O}(\partial^2)$ contributions (or the more recently studied $\mathcal{O}(\partial^4)$ ones). See, however, Ref.~\cite{PhysRevB.94.140506} where the order $\mathcal{O}(\partial^2)$ was implemented in order to study the critical Casimir forces. Moreover, they have been done before the recent developments concerning the convergence of the DE and, as a consequence, do not include estimates of error bars. The purpose of the present article is to reconsider the calculation of UARs within the DE of the NPRG including the up-to-date developments.

We present in the next section a brief introduction to the NPRG and the DE. In Section \ref{UARdef} we review the definition and main characteristics of UARs. Then, we present our main results in Section \ref{results} and, finally, we discuss our conclusions. More technical material concerning numerical and procedural details is presented in appendices.

\section{Universal amplitude ratios: definitions and examples}\label{UARdef}

In this section we discuss the key ideas behind the universality of UARs, introduce those that will be calculated in this work and briefly review earlier results in the literature. 

\subsection{Brief review on the origin of universality}

Just like critical exponents, UARs are --as their name implies-- universal, in the sense that their values do not depend on the details of the microscopic physics, but rather on general properties of the model under consideration such as the dimension of space, the symmetry group of the model and the representation of this group spanned by the order parameter. Most of the presented material in this subsection is elementary but, nevertheless, we included it for notational purposes and to make the presentation self-contained. Our presentation follows classical references on the topic, such as \cite{Pelissetto02,ZinnJustin:2002ru,privman1991universal}.

As is well known, in the models we consider, corresponding to the $O(N)$ universality class, critical phenomena are characterized by the fact that the RG flow possesses a non-trivial fixed point, known as the Wilson-Fisher (WF) point. Crucially, the basin of attraction of this fixed point has co-dimension two\footnote{As explained in subsequent paragraphs, these correspond to a $Z_2$ even perturbation (usually associated with the temperature) and a $Z_2$ odd perturbation (usually associated with the external magnetic field).} in the space of coupling constants. Let us parametrize the couplings $\{u_i\}_{i>0}$ in such a way that close to the WF fixed point $u_1$ and $u_2$ coincide with the two unstable eigendirections of the linearized RG flow.\footnote{Notice that we are assuming that there are no marginal operators in the spectrum of the fixed point and thus the codimension of its basin of attraction coincides with the number of its unstable eigendirections.} For sufficiently small but non-zero control parameters $t$ and/or $H$, the resulting RG orbit passes sufficiently close to the fixed point where $t=0$ and $H=0$ and departs along some unstable trajectory. While this unstable trajectory is close to the fixed point, it is well approximated by the linearized flow, which, with an appropriate parametrization of coupling constants, is given by, to leading order in $H$ and $t$,
\begin{equation}\label{linearflowui}
\begin{split}
u_1 &\approx c_1 t e^{y_1 s},\\
u_2 &\approx c_2 H e^{y_2 s},\\
u_i &\approx b_i e^{y_i s}\hspace{4pt}i>2,
\end{split}
\end{equation}
for some constants $c_1$, $c_2$ and $b_i$ which depend on the details of how the two-parameter family of initial conditions is defined, the $b_i$ being generically independent of $t$ and $H$ to leading order in these quantities. The $c_i$ and $b_i$ coefficients carry the information of the particular microscopic theories defining our family of RG flows. Here $s$ is the RG ``time'' and the $y_i$ are the eigenvalues of the linearized flow. The unstable eigenvalues $y_1$ and $y_2$ are negative and all other eigenvalues are positive. These two parameters are interpreted, in the case of magnetic systems, as the reduced temperature $t = (T-T_c)/T_c$ and the external magnetic field strength $H_i = Hn_i$, $n_in_i =1$ respectively.

To study the behavior of macroscopic quantities for these theories, we need to examine how the thermodynamic functions evolve under the flows described in the previous paragraph. Let us consider for instance the Gibbs free energy $\mathcal{F}$ as a function of the parameters $u_i$. For the RG flows we are considering, it can be seen that, in the scaling region where Eq.~\eqref{linearflowui} is valid, the singular part of $\mathcal{F}$ flows according to
\begin{equation}\label{scalingGibbs}
\begin{split}
\mathcal{F}_{sing}&(u_1,u_2,\dots,u_{i>2},\dots) \approx\\ &\lambda^{-d}\mathcal{F}_{sing}(c_1 t \lambda^{y_1},c_2 H \lambda^{y_2},\left\{b_i\lambda^{y_i}\right\}_{i>2}),
\end{split}
\end{equation}
where $\lambda\equiv e^{s}$ and $d$ is the dimension of space. The crucial aspect of this formula as far as universality is concerned is that the RHS has no implicit dependence on $\lambda$ but rather all dependence on the RG time is in the evolution of the $u_i$ as shown. Indeed, all of the usual statements about universality follow from this fact.

\subsection{High-temperature universal equation of state}

To see more precisely how the aforementioned universal behavior comes about let us for now restrict our discussion to the high temperature phase, where $t>0$. Then, if we evaluate the RHS of Eq.~\eqref{scalingGibbs} at $\lambda = (c_1 t)^{-1/y_1}$ we get
\begin{equation}\label{gibbsUniv}
\mathcal{F}_{sing} = (c_1 t)^{\frac{d}{y_1}}\mathcal{F}_+(y),
\end{equation}
where we have defined
\begin{equation}
\mathcal{F}_+(y) \equiv \mathcal{F}_{sing}(1,y,0,\dots,0),
\end{equation}
with
\begin{equation}
y \equiv c_2(c_1 t)^{-\frac{y_2}{y_1}}H.
\end{equation}

Crucially we have assumed that the value of $\lambda$ at which we have evaluated the RHS of Eq.~\eqref{scalingGibbs} is such that the dependence on the $u_{i>2}$ coupling constants can be neglected. Notice that the previous statement about the RHS of Eq.~\eqref{scalingGibbs} not depending implicitly on $\lambda$ beyond the dependences shown in Eq.~\eqref{scalingGibbs} implies that the function $\mathcal{F}_+(y)$ is universal. From  Eq.~\eqref{scalingGibbs} we may calculate the susceptibilities at zero external field
\begin{equation}
\chi_{2n} \equiv -\frac{\partial^{2n}\mathcal{F}}{(\partial H_i\partial H_i)^n}\bigg\rvert_{H=0},
\end{equation}
and find\footnote{To get the factors correctly the relation $(\partial_H\cdot \partial_H)^n(H\cdot H)^n= N(N+2)\dots(N+2n-2)2^n n!$ is useful.}
\begin{equation}\label{suscept}
\chi_{2n} = -c_2(c_1)^{-\gamma_{2n}}R_{n,N}\mathcal{F}_{+}^{(2n)}t^{-\gamma_{2n}},
\end{equation}
where
\begin{align}\label{gamman_from_lineariz}
R_{n,N} &\equiv N(N+2)\dots(N+2n-2)/(2n-1)!!, \\
\mathcal{F}_{+}^{(2n)} &\equiv \partial^{(2n)}_y\mathcal{F_+}\vert_{y=0},\\
\gamma_{2n} &\equiv \frac{y_2}{y_1}2n -\frac{d}{y_1}.
\end{align}

Before we proceed further let us make a small aside to point out an important subtlety in the procedure just described. In writing Eq.~\eqref{linearflowui} we have mentioned that this holds for the linearized flow. However, taking $\lambda = (c_1 t)^{-1/y_1}$ would appear to correspond, for small values of $t$, to choosing a value of $s$ which may be outside of the region where the linearization of the flow is valid. The way to take care of this apparent contradiction is to consider a recoordinatization of the space of coupling constants such that the flow has an exponential form similar to Eq.~\eqref{linearflowui} in a finite region around the WF fixed point (see for instance \cite{wegner1972corrections}). This amounts to a modification of \eqref{linearflowui} which ends up affecting the scaling relations only through corrections to the leading scaling behavior which do not affect the conclusions that follow (see \cite{Pelissetto02} for details).

In Eq.~\eqref{suscept}\footnote{It is important to point out that what is shown in Eq.~\eqref{suscept} is in fact the behavior at leading order in $t$; there will be corrections although they will not concern us here.} the factor multiplying $t^{-\gamma_{2n}}$ is clearly not universal since it depends on $c_1$ and $c_2$. However, it is proportional to the $\mathcal{F}_+^{(2n)}$ which are indeed universal, since they are extracted from the function $\mathcal{F}_+(x)$ and, in fact, it is possible to determine this universal information through appropriate quotients of physical quantities. In order to do this it is useful to note that the two point correlation function satisfies a similar scaling equation to Eq.~\eqref{scalingGibbs} and that from such an equation it is straightforward to obtain that, for small positive $t$, the correlation length behaves as
\begin{equation}\label{scaling_xi}
\xi = c_1^{-\nu}t^{-\nu},
\end{equation}
where
\begin{equation}\label{nu_from_lineariz}
\nu = -\frac{1}{y_1}.
\end{equation}    

We now review how some standard results about universality are obtained. First, from Eq.~\eqref{gamman_from_lineariz} and Eq.~\eqref{nu_from_lineariz} we see that the critical exponents $\gamma_{2n}$ and $\nu$ satisfy the hyperscaling relations
\begin{equation}\label{hyperscaling_gamma}
\gamma_{2n} = \gamma_{2}n + \nu d (n-1).
\end{equation} 

Similar arguments using also Eq.~\eqref{suscept} and Eq.~\eqref{scaling_xi} allow to write a quotient of universal quantities $\mathcal{F}^{(2n)}_+$ in terms of appropriate ratios of powers of the susceptibilities and correlation length
\begin{equation}\label{firstUAR}
\frac{\mathcal{F}^{(2n)}_+}{(\mathcal{F}^{(2)}_+)^n} = \frac{\chi_{2n}}{(\chi_2)^{n}\xi^{d(n-1)}}\frac{R_{1,N}^n}{R_{n,N}}.
\end{equation}

Notice that the hyperscaling relations Eq.~\eqref{hyperscaling_gamma} imply that the RHS of this equation is $t$ independent to leading order in $t$. Just to connect with the usual expressions found in the literature, we note that if the scaling forms of $\chi_{2n}$ and $\xi$ to leading order in $t$ are written as
\begin{align}\label{scalingchi}
\chi_{2n} = R_{n,N}C^{+}_{2n}t^{-\gamma_{2n}},\\
\xi = f^+t^{-\nu},
\end{align}
then the ratio in \eqref{firstUAR} is usually written as
\begin{equation}
\frac{\mathcal{F}^{(2n)}_+}{(\mathcal{F}^{(2)}_+)^n} =\frac{C^{+}_{2n}}{(C^{+}_{2})^n(f^+)^{d(n-1)}}.
\end{equation}

The quantities given by Eq.~\eqref{firstUAR} are our first example of UARs: a ratio of --non-universal-- physical quantities that yields a universal result. These particular UARs characterize universal features of the high temperature phase of the $O(N)$ models. In particular, one such quantity which will be calculated in this paper is

\begin{equation}\label{eq:g_4FromFs}
g_4 \equiv -\frac{\mathcal{F}^{(4)}_+}{(\mathcal{F}^{(2)}_+)^2},
\end{equation}
where the extra minus sign is conventional. For higher values of $n$ the corresponding ratios are not usually discussed in the literature but rather combinations of them which are obtained as follows. We first write the magnetization at nonzero external field $H$ in terms of the Gibbs free energy to get
\begin{equation}
M_i = -\partial_{H^i}\mathcal{F}_{sing}=-c_2(c_1t)^{\frac{d-y_2}{y_1}} \partial_yF_+(y)\frac{H_i}{H}.
\end{equation}
Using that in the high temperature phase the expansion of $\mathcal{F_+}$ around zero only involves even powers of $y$ it is straightforward to invert the previous relation to obtain
\begin{equation}\label{highTEOS}
H_i  = \frac{M_i}{M}c_2^{-1}(c_1)^{\frac{y_2}{y_1}}\left(\frac{-\mathcal{F}^{(4)}_+}{\mathcal{F}^{(2)}_+}\right)^{-1/2}t^{\frac{y_2}{y_1}}G(z),
\end{equation}
where
\begin{equation}
z \equiv M c_2^{-1}c_1^{\frac{y_2-d}{y_1}}\left(\frac{-\mathcal{F}^{(4)}_+}{(\mathcal{F}^{(2)})^3}\right)t^{\frac{y_2-d}{y_1}}.
\end{equation}
The factors in the RHS of Eq.~\eqref{highTEOS} are chosen in such a way that the expansion of $G(z)$ around zero is of the form
\begin{equation}
G(z) = z + \frac{z^3}{3!} + \sum_{n>1}r_{2n+2}
\frac{z^{2n+1}}{(2n+1)!}.
\end{equation}
The function $G(z)$ is universal as its coefficients $r_n$ are easily obtainable in terms of the $\mathcal{F}_+^{(2n)}$. For instance, we have
\begin{align}
r_6 &= 10 - \frac{\mathcal{F}^{(2)}_+\mathcal{F}^{(6)}_+}{(\mathcal{F}^{(4)}_+)^2},\\
r_8 &= 280 -56\frac{\mathcal{F}^{(2)}_+\mathcal{F}^{(6)}_+}{(\mathcal{F}^{(4)}_+)^2} + \frac{(\mathcal{F}^{(2)}_+)^2\mathcal{F}^{(8)}_+}{(\mathcal{F}^{(4)}_+)^3},\\
r_{10} &=   15400 -4620\frac{\mathcal{F}^{(2)}_+\mathcal{F}^{(6)}_+}{(\mathcal{F}^{(4)}_+)^2} + 126\left(\frac{\mathcal{F}^{(2)}_+\mathcal{F}^{(6)}_+}{(\mathcal{F}^{(4)}_+)^2}\right)^2\nonumber\\ 
&+120 \frac{(\mathcal{F}^{(2)}_+)^2\mathcal{F}^{(8)}_+}{(\mathcal{F}^{(4)}_+)^3}-\frac{(\mathcal{F}^{(2)}_+)^3 \mathcal{F}^{(10)}_+}{(\mathcal{F}^{(4)}_+)^4}.
\end{align}
Using Eq.~\eqref{scalingchi} and Eq.~\eqref{suscept} they can be written in terms of ratios of the $C^+_{2n}$ as
\begin{align}
r_6 &= 10 -\frac{C_2^+C_6^+}{(C_4^+)^2},\\ 
r_8 &= 280 -56\frac{C_2^+C_6^+}{(C_4^+)^2} + \frac{(C_2^+)^2C_8^+}{(C_4^+)^3},\\
r_{10} &= 15400 -4620\frac{C_2^+C_6^+}{(C_4^+)^2}+126\left(\frac{C_2^+C_6^+}{(C_4^+)^2}\right)^2\nonumber\\
&+ 120  \frac{(C_2^+)^2C_8^+}{(C_4^+)^3} - \frac{(C_2^+)^2 C_{10}^+}{(C_4^+)^4},
\end{align}
which we reproduce here for completeness as it is in this form that these quantities are often found in the literature.

\subsection{UARs beyond the high temperature regime}

So far, we have only discussed UARs that are built out of ratios of quantities obtained in the high temperature phase. However, through a similar reasoning to that we have presented so far it can be seen that there are UARs that can be obtained by combining quantities corresponding to the high temperature phase, the low temperature phase and at the critical temperature. We will not discuss them in the same level of detail as presented so far for the $g_4$ and $r_i$ but merely give the definitions of those that will be calculated in this work. In order to do so, we first review the leading scaling behaviors in each of the phases of some of the quantities involved in these definitions, mainly to set up notations for the definitions of the UARs.

In the high temperature phase at zero external field, aside from the quantities discussed so far, we also have the leading scaling behavior at $t\rightarrow0^+$ of the specific heat, which is given by
\begin{equation}\label{hightscalingspecificheat}
T_cC_H = A^+t^{-\alpha}.
\end{equation}

In the low temperature phase at zero external field the system presents a nonzero magnetization, whose leading scaling behavior with the reduced temperature $t$ is written as
\begin{equation}\label{scalingmag}
M = B(-t)^{\beta}
\end{equation} 
In this phase the specific heat presents, to leading order in $t$, a power law behavior similar to Eq.~\eqref{hightscalingspecificheat} which we write as
\begin{equation}\label{lowtscalingspecificheat}
T_cC_H = A^-t^{-\alpha}.
\end{equation}

We also consider the system at the critical temperature $t=0$ but nonzero external field $H$, what is known as the critical isotherm. Here the magnetization presents at leading order in the external field a power law behavior which we write as

\begin{equation}
M=B_cH^{1/\delta}.
\end{equation}

Having fixed these notations, the following quantities are universal in the sense previously described and for the same reasons:
\begin{align}
U_0 &\equiv \frac{A^+}{A^-},\\
R_{\chi} &\equiv \frac{C_2^+ B^{\delta-1}}{B_c^{\delta}},\\
R_c &\equiv \frac{\alpha A^+C_2^+}{B^2},\\
R_4 &\equiv -\frac{C_4^+B^2}{(C_2^+)^3},\\
R_{\alpha} &\equiv \frac{1-U_{0}}{\alpha},\\
R_{\xi}^+ &\equiv \left(\frac{R_cR_4}{g_4}\right)^{1/3}.
\end{align}
Notice that the last two quantities are not independent of the other UARs. We introduce the definition of these for completeness, given that in the literature results are often presented in terms of them.

Finally, some UARs involve quantities calculated at what is known as the crossover line, which is by definition the value of the reduced temperature $t_m$ for which the longitudinal susceptibility $\chi_L$ is maximum at fixed external field $H$. Here the scaling laws are
\begin{align}
t_m =	T_p H^{1/(\gamma+\beta)},\\
\chi_L = C_p t_m^{-\gamma}.
\end{align} 
In these terms one may define:
\begin{align}
P_m &\equiv \frac{T_p^{\beta}B}{B_c},\\
R_p &\equiv \frac{C^+_{2}}{C_p},\\
P_c &\equiv \frac{P_m^{2\delta}}{R_{\chi}^2R_4},
\end{align}
which complete the list of UARs that will be calculated in this work. Notice that once again these quantities are not independent, as $P_c$ is defined exclusively in terms of other UARs.

The universal quantities that we have described in this section have been previously calculated for $O(N)$ models through a variety of methods such as the $\epsilon$-expansion, the high-temperature expansion or Monte Carlo methods. In the next section we present a summary of some of the previous results that can be found in the literature for the most commonly studied values of $N$. We refer the reader to \cite{Pelissetto02} for a more exhaustive list of results.

\section{Non-perturbative Renormalization Group and the Derivative Expansion}

We give in this section a short review of the NPRG method and of the approximation scheme to be employed in the present article, namely the Derivative Expansion. The main purpose here is to fix notations. A recent detailed review of these topics can be found in \cite{Dupuis:2020fhh}. A detailed analysis of the DE in the $O(N)$ models, where the convergence of the method and a procedure to estimate error bars has been studied, can be found in \cite{Balog:2019rrg,DePolsi:2020pjk}. We follow here the main lines of presentation of \cite{DePolsi:2020pjk} for completeness purposes but we refer the reader to \cite{Balog:2019rrg,DePolsi:2020pjk,Dupuis:2020fhh} for a more detailed discussion.

\label{nprgandde}
\subsection{The Non-Perturbative Renormalization Group}

The NPRG is based on Wilson's ideas of integrating over modes with a wave-number larger than some scale $k$ while keeping the long-distance modes frozen. This is done by introducing an infrared regulator in the theory.

In order to do so we add to the Hamiltonian a quadratic term in the fields \cite{Polchinski:1983gv}, $S[\varphi]\to S[\varphi]+\Delta S_k[\varphi]$ with:
\begin{equation} \label{deltaS}
\Delta S_k[\varphi]=\frac 1 2 \int_{q}\varphi_a(-q) R_k(q^2) \varphi_a(q),
\end{equation}
where $\int_q=\int \frac{d^dq}{(2\pi)^d}$. Here and below, the Einstein convention is employed for sums over internal indices. To properly act as an infrared regulator, $R_k(q^2)$ should:
\begin{itemize}
	\item be a smooth function of the momentum squared $q^2$;
	\item $R_k(q)\sim Z_k k^2$ for $q\ll k$, where $Z_k$ is a field renormalization factor to be specified below;
	\item $R_k(q)\to 0$ faster than any power law when $q\gg k$.
\end{itemize}

One can then define a scale-dependent free-energy $W_k[J]$ \cite{Wetterich:1992yh,Ellwanger:1993kk,Morris:1993qb}:
\begin{equation} \label{regulatedgeneratingfunc}
e^{W_k[J]}=\int \mathcal{D}\varphi\  e^{-S[\varphi]-\Delta 
	S_k[\varphi]+\int_x J_a(x) \varphi_a(x)},
\end{equation}
where $\int_x=\int d^dx$.
The scale-dependent effective action $\Gamma_k[\phi]$, is defined as the modified Legendre transform of $W_k[J]$:
\begin{equation}
\label{legendre}
\Gamma_k[\phi]=\int_x \phi_a(x) J^\phi_a(x) -W_k[J^\phi]-\Delta S_k[\phi],
\end{equation}
where, $J^\phi$ is an implicit function of $\phi$, obtained by inverting
\begin{equation}
\phi_a(x)=\frac{\delta W_k}{\delta J_a(x)}\Bigg|_{J=J^\phi}.
\end{equation}
From the properties of the regulator $R_k$ listed above and Eq.~(\ref{legendre}), it can then be shown that at scale $k=\Lambda$, $\Gamma_\Lambda[\phi]\sim S[\phi]$. This will provide the initial condition of the exact RG flow given below in Eq.~(\ref{wettericheq}).

$\Gamma_k[\phi]$ is the generating functional of infrared-regularized one-particle irreducible (1PI) correlation functions or proper vertices (that we choose to evaluate at a uniform field). Its Fourier transform is defined as:
\begin{align} \label{eqgamma_n}
\Gamma^{(n)}_{a_1\dots a_n}(p_1,\dots,&p_{n-1};\phi)=\int_x \,\mathrm{e}^{i \sum_{m=1}^{n-1}x_m\cdot 
	p_m} \nonumber \\
&\times\Gamma^{(n)}_{a_1\dots a_n}(x_1,\dots,x_{n-1},0;\phi).
\end{align}
where, due to translational invariance, $\Gamma^{(n)}$ only depends on $n-1$ independent wave-vectors. 

The evolution of $\Gamma_k[\phi]$ with the RG time $s=\log(k/\Lambda)$ \cite{Wetterich:1992yh,Ellwanger:1993kk,Morris:1993qb} can be easily obtained:
\begin{equation}\label{wettericheq}
\partial_{s}\Gamma_{k}[\phi]=\frac{1}{2}\int_{x,y}\partial_{s}R_{k}(x-y)G_{aa}[x,y;\phi].
\end{equation}
Here $R_{k}(x-y)$ is the Fourier transform of $R_{k}(q^2)$ and $G_{ab}[x,y;\phi]$ is the dressed propagator in an arbitrary external field $\phi(x)$. The latter can be obtained from the two-point vertex by
\begin{equation}
\int_{y} G_{ac}[x,y;\phi]\Big[\frac{\delta^2\Gamma_k[\phi]}{\delta \phi_c(y)\delta \phi_b(z)}+ R_k(y-z)\delta_{cb}\Big]=\delta(x-z)\delta_{ab}.
\end{equation}
We omit the $k$-dependence of the propagator and proper vertices to alleviate the notation. Taking successive functional derivatives of Eq.~\eqref{wettericheq}, one can derive equations for all proper vertices.  For instance, evaluating Eq.~\eqref{wettericheq} in a uniform external field one deduces the exact equation for the effective potential $U_k$ (or ``0-point vertex''):
\begin{equation}
\label{eqpot}
\partial_s U_k(\rho)=\frac 1 2 \int_q \partial_s R_k(q) G_{aa}(q;\phi),
\end{equation}
where $\rho=\phi_a\phi_a/2$ and $G_{ab}(q;\phi)$ is the Fourier transform of the propagator evaluated in a uniform field. In the same way, the equation for the 2-point function in a uniform external field can be deduced, yielding:
\begin{align}
\partial_s &\Gamma_{ab}^{(2)}(p;\phi)=\int_q \partial_s R_k(q) G_{mn}(q;\phi)
\Big\{-\frac 1 2 \Gamma_{abno}^{(4)}(p,-p,q;\phi)\nonumber\\
&+ \Gamma_{anl}^{(3)}(p,q;\phi)G_{lr}(p+q;\phi)\Gamma_{bro}^{(3)}(-p,-q;\phi)
\Big\}G_{om}(q;\phi).
\label{eqvertex2}
\end{align}
The equation for a given vertex $\Gamma^{(n)}$ depends on all the vertices up to $\Gamma^{(n+2)}$. As a consequence, one has an infinite hierarchy of coupled equations for the vertex functions. Solving this equation typically requires approximations.

The advantage of Eq.~\eqref{wettericheq} with respect to other field-theoretical approaches is that it is well-suited to formulate approximations going beyond perturbation theory. We now present the most employed approximation in the NPRG context: the DE. 

\subsection{The Derivative Expansion}
\label{DEgeneral}
The DE procedure corresponds to expanding all proper vertices in power series of the momenta to a finite order (except for the 2-point function where, at order zero, one keeps the bare momentum dependence in the propagator). As a consequence, this approximation is only valid for the low-momentum physics. In fact, this has proven to be a very precise approximation scheme in three dimensions for the calculation of critical exponents in the $\mathbb{Z}_2$ and $O(N)$ universality classes (see for example, \cite{Balog:2019rrg,DePolsi:2020pjk,Peli:2020yiz,Dupuis:2020fhh}). In the past, various reasons for the success of the DE in $O(N)$ models have been proposed in the literature \cite{DAttanasio:1997yph,Berges:2000ew,Canet:2003qd,Canet:2002gs,Blaizot:2005xy,Delamotte:2007pf,Benitez:2011xx}. A first reason is that integrals in equations for proper vertices such as Eq.~\eqref{eqpot}) or Eq.~\eqref{eqvertex2} include the derivative of the regulating function $\partial_s R_k(q)$ in the numerator. The integral over $q$ is then dominated by the range $q \lesssim k$. As a consequence, expanding in all momenta (including the internal one) gives equations that couple only weakly to the large momentum sector $p\gg k$. This allows for the formulation of the DE approximation scheme for momenta smaller than the maximum between $k$ and the inverse of the correlation length. For critical phenomena, when $k\to 0$ the DE only applies to those quantities dominated by zero momenta (such as thermodynamic properties or critical exponents). Even if the expansion makes sense at low momenta, it had been unclear, until recently, why it should give good results because {\it a priori} the associated expansion parameter would be $q^2/k^2$, which reaches values of order one because the derivative of the regulator in equations such as (\ref{eqpot}) or (\ref{eqvertex2}) suppresses all momenta  beyond typically $k$. This is the main reason why the DE has been seen for a long time as an ``uncontrolled'' approximation. 

This changed recently with some studies on the radius of convergence of the DE \cite{Balog:2019rrg,DePolsi:2020pjk,Dupuis:2020fhh}\footnote{It is interesting to note that the convergence of the DE was studied long time ago in the perturbative regime with similar results to those reviewed here \cite{Morris:1999ba}.}. Although there is some dependence on the model considered and on the chosen regulating function $R_k$, the convergence is governed by very general features as was shown in \cite{Balog:2019rrg}. To be precise, in models described by Ginzburg-Landau Hamiltonians whose analytical continuation to the Minkowskian space gives unitary models, the radius of convergence has been shown to be of the order  $q_{radius}^2/k^2\simeq 4-9$ \cite{Balog:2019rrg} if a reasonable regulator is chosen. For this to hold, one needs the theory to behave, for momenta $q^2/k^2\lesssim 4-9$, as a massive theory with a mass of order $k$ and that, at the same time, momenta $q>k$ be strongly suppressed in flow equations.
In the present article, we will employ two of the regulators employed in \cite{DePolsi:2020pjk} which were shown to fulfil these requirements:
\begin{subequations}
	\begin{align}
	W_k(q^2) &= \alpha Z_k k^2 \, y/(\exp(y) - 1), \label{regulator-wetterich}\\ 
	E_{k}(q^2) &= \alpha Z_k k^2 \,\exp(-y).  \label{regulator-exp}
	\end{align}
	\label{regulators}
\end{subequations}
We will test the dependence of the results by varying
$\alpha$ for each family of regulators given in (\ref{regulator-wetterich}) and (\ref{regulator-exp}). The optimal value of $\alpha$ is determined by imposing the ``Principle of  Minimal Sensitivity'' (PMS) \cite{Stevenson:1981vj,Canet:2002gs}.
Given the fact that without approximations physical predictions should not depend on the regulator (and, in particular, on $\alpha$), we consider as optimum the value of $\alpha$ corresponding to an extremum.\footnote{Very recently, it was shown that, at least in the Ising universality class, the PMS can be justified by requiring the fixed point obtained from the DE to be as conformal as possible \cite{Balog:2020fyt}.} By employing this procedure and for momenta below $k$, successive corrections in the DE for critical exponents are suppressed by a factor $1/9$ -- $1/4$. As will be shown in the present article the same procedure works as well for UARs. As a final remark, we mention that the quality of most DE results is further improved in all cases where the exponent $\eta$ is small because, as explained in Ref.~\cite{Balog:2019rrg}, all subleading orders in the flow of 2-point functions are suppressed by a factor of $\eta$.

All the previous analysis about the convergence of the DE is applicable for $\mathbb{Z}_2$ and $O(N)$ Ginzburg-Ladau models with $N\geq 1$. This has been confirmed by specific calculations of critical exponents in the Ising universality class at various orders up to order $\mathcal{O}(\partial^6)$ \cite{Canet:2003qd,Balog:2019rrg} for the Ising universality class, and up to order $\mathcal{O}(\partial^4)$ for $O(N)$ models \cite{DePolsi:2020pjk,Peli:2020yiz}.\footnote{It is important to stress that the radius of convergence of order $q^2/k^2\sim 4$--$9$ only applies for an equation such as Eq.~\eqref{wettericheq} where only 1PI contributions are present. The DE implemented in FRG equations based on 1-particle reducible correlation functions such as the Wilson-Polchinski equation  \cite{Wilson:1973jj,Polchinski:1983gv} has a radius of convergence of order $q^2/k^2 \sim 1$. This explains why the DE gives much better results in the NPRG formulation than in Wilson-Polchinski one \cite{Bervillier:2005za,Morris:1999ba}.}

Previous analysis of the convergence of the DE focus on critical exponents. However, essentially all the previous discussion about the convergence of the DE applies to any observable quantity dominated by momenta smaller than the inverse of the correlation length. In the particular case of the critical regime, this reduces to thermodynamical properties, or other quantities such as critical exponents that can be extracted from vertices or their derivatives evaluated at zero momenta (whether they be universal or not). In the present article we want to generalize the results of Refs.~\cite{Balog:2019rrg,DePolsi:2020pjk} for critical exponents to other quantities. We will nevertheless focus in the following on the UARs for three reasons.
First, these quantities have a wide theoretical and experimental interest. Second, they have been largely studied by using a variety of methods and are known with relatively good precision. This makes them excellent benchmarks for testing approximation schemes. Third, in spite of being universal as explained in the introduction, UARs do not depend only on the critical fixed point but require the exploration of some aspects of the low and high temperature phases. As a consequence, their direct extraction from methods based on conformal symmetry is very hard (or may be impossible). As will be discussed in the next section, UARs can be extracted from vertices and their derivatives at zero momenta. This is why they are in the domain of validity of the DE.

Focusing on universal quantities, one can choose any reasonable microscopic Hamiltonian in a given universality class and so we will employ for simplicity a Ginzburg-Landau model with Hamiltonian,
\begin{equation}\label{GLmodel}
S[\varphi]=\int_x \Big\{\frac{1}{2}\big(\partial_\mu\varphi^a\big)^2+\frac r 2 
\varphi^a\varphi^a+
\frac{u}{4!}(\varphi^a\varphi^a)^2\Big\}.
\end{equation}
One then implements the DE. That is, one considers for $\Gamma_k$ the most general terms compatible with the symmetries of a given universality class with, at most, a given number of derivatives. Linear symmetries of the Hamiltonian (\ref{GLmodel}) that are preserved by the regulator $\Delta S_k[\phi]$, are ensured along the flow. That is, $\Gamma_k[\phi]$ must satisfy space isometries and internal $O(N)$ symmetry.

At order $\mathcal{O}(\partial^4)$, this gives the \textit{ansatz} \cite{DePolsi:2020pjk}:
\begin{align}\label{ansatz-order4}
\nonumber\Gamma_k^{\partial^4} &[\phi]=\int_x \Big\{  
U_k(\rho)+\frac{1}{2}Z_k(\rho)\big(\partial_\mu\phi^a\big)^2
+\frac{1}{4}Y_k(\rho)\big(\partial_\mu\rho\big)^2
\\	\nonumber
&+\frac{W_1(\rho)}{2}\big(\partial_\mu\partial_\nu \phi^a\big)^2  \nonumber
+\frac{W_2(\rho)}{2}\big(\phi^a \partial_\mu\partial_\nu \phi^a\big)^2 \\
& \nonumber + W_3(\rho)\partial_\mu\rho\partial_\nu\phi^a\partial_\mu\partial_\nu \phi^a
+\frac{W_4(\rho)}{2}
\phi^b\partial_\mu\phi^a\partial_\nu\phi^a\partial_\mu\partial_\nu \phi^b \\
\nonumber &
+ \frac{W_5(\rho)}{2}
\varphi^a\partial_\mu\rho\partial_\nu\rho\partial_\mu\partial_\nu 
\varphi^a
+\frac{W_6(\rho)}{4}\Big(\big(\partial_\mu\varphi^a\big)^2\Big)^2 \\
\nonumber & +\frac{W_7(\rho)}{4}\big(\partial_\mu\phi^a\partial_\nu\phi^a\big)^2
+\frac{W_8(\rho)}{2}
\partial_\mu\phi^a\partial_\nu\varphi^a\partial_\mu\rho\partial_\nu\rho
\\ &
+\frac{W_9(\rho)}{2}\big(\partial_\mu\varphi^a\big)^2 \big(\partial_\nu\rho\big)^2 
+\frac{W_{10}(\rho)}{4} \Big(\big(\partial_\mu \rho)^2\Big)^2  \Big\}.
\end{align}
In the previous expression, for notational simplicity, we omitted the $k$ dependence in the W's. 
Notice that for $N=1$, some terms are not independent of each other. This is for instance the case of the $Z_k(\rho)$ and $Y_k(\rho)$ terms: Including them in this case would therefore be redundant. A direct inspection shows that only three terms  at order $\mathcal{O}(\partial^4)$ are independent \cite{Canet:2003qd} when $N=1$ and therefore need to be included in the ansatz (\ref{ansatz-order4}).

The flow of the various functions is implemented in Fourier space. 
For instance, the flow of the effective potential $U_k(\rho)$ is obtained at order $\mathcal{O}(\partial^4)$ of the DE from Eq.~(\ref{eqpot}) by inserting the propagator $G_k(q;\phi)$ computed from $\Gamma^{(2)}_k(q;\phi)$ which is itself obtained from the {\it ansatz} (\ref{ansatz-order4}):
\begin{align}\label{eqgamma2DE2}
\Gamma_{ab}^{(2)}(p;\phi)
&=\delta_{ab}\big(U_k'(\rho)+Z_k(\rho)p^2+W_1(\rho)p^4\big)\nonumber\\
&+\phi_a \phi_b \big(U_k''(\rho)+\frac{1}{2}Y_k(\rho)p^2+W_2(\rho)p^4\big)+\mathcal{O}(p^6).
\end{align}
Similarly, the equation for $Z_k(\rho)$, $Y_k(\rho)$, $W_1(\rho)$ or $W_2(\rho)$ can be obtained from the equation for the 2-point function (in a uniform external field). For that purpose, one expresses those functions in terms of the vertices. For example,
\begin{equation}
Z_k(\rho)=\frac{1}{N-1}\Big(\delta_{ab}-\frac{\phi_a\phi_b}{2\rho}\Big)\partial_
{p^2}\Gamma_{ab}^{(2)}(p;\phi)\vert_{{\bf p}=0}.
\end{equation}
We employed the flow equations obtained previously for arbitrary $N$ at order $\mathcal{O}(\partial^4)$ in Ref.~\cite{DePolsi:2020pjk}. We want to point out, however, that we considered, as in previous references \cite{Balog:2019rrg,DePolsi:2020pjk}, a strict polynomial expansion in momenta at the order of the DE being considered in the product of vertices (see \cite{DePolsi:2020pjk} for details). This differs from more standard implementations of the DE \cite{VonGersdorff:2000kp,Canet:2003qd,Peli:2020yiz}.\footnote{Both versions of the DE give results for critical exponents that are compatible within error bars at order $O(\partial^2)$ \cite{DePolsi:2020pjk}.} The details of the numerical solving of the equations is presented in Appendix~\ref{Ap:NumParam}.

In the present work we are interested in quantities that are dominated by the neighbourhood of the critical point. In order to reach such point we need to fine-tune one bare symmetric coupling in the initial condition for the flow equations.

To identify the critical point one can exploit the fact that it is scale invariant. The Ward identities for scale invariance in presence of the infrared regulator $\Delta S_k$ are equivalent to the fixed point condition $\partial_s\Gamma_k=0$ when $\Gamma_k$ is expressed in terms of dimensionless and renormalized quantities \cite{delamotte2016scale}. To be precise, one defines renormalized and dimensionless fields and coordinates by
\begin{align}
\tilde x&=k x,\\
\phi^a(x)&=k^{(d-2)/2}Z_k^{-1/2}\tilde \phi^a(\tilde x), \\
\rho(x)&=k^{(d-2)}Z_k^{-1}\tilde \rho(\tilde x),
\end{align}
and functions $\tilde{F}(\tilde\rho(\tilde{x}))$:
\begin{equation}
F(\rho)=k^{d_F}Z_k^{n/2}\tilde{F}(\tilde\rho),
\end{equation}
where $F(\rho)$ is any function involved in the {\it ansatz} (\ref{ansatz-order4}) for $\Gamma_k$, $d_F$ is the canonical dimension of $F$ and $n$ the number of fields $\phi^a$ that multiply $F$ in $\Gamma_k$. As for $Z_k$ (not to be confused with the function $Z_k(\rho)$) it is given by $Z_k(\rho)=Z_k\tilde{Z}_k(\tilde\rho)$ once a normalization condition is specified. We use the normalization condition: $\tilde Z_k(\tilde\rho_0)=1$ for a fixed value of $\tilde\rho_0$. The scale-dependent anomalous dimension is $\eta_k= -\partial_s\log Z_k$ which, at the fixed point, becomes the physical anomalous dimension $\eta$ \cite{Berges:2000ew}.

Once the critical point has been reached, to study the UARs one must typically de-tune the bare parameters of the model in order to go to the high or low temperature phases, as is discussed in subsequent sections.

\subsection{Central values and error bar estimates}\label{errorbars}

One of the main advantages of knowing the existence of a ``small parameter'' (of about $1/9$--$1/4$) controlling successive orders of the DE is that one is able to estimate error bars for the predictions made with this approximation scheme (at least for models where a unitary Minkowskian extension exists). The strategy to be employed in the present work to estimate error bars was presented first in detail in \cite{DePolsi:2020pjk}. For completeness we recall the most relevant elements here.

Let's consider a physical quantity $Q$ such as a critical exponent or a UAR dominated by momenta smaller than or of the order of the inverse of the correlation length. To estimate an error bar for the calculation of this quantity we proceed as follows:
\begin{itemize}
	\item One calculates the quantity $Q$ by using various families of regulators. Each family is parametrized by the multiplicative factor $\alpha$ introduced for each regulator family \cite{Canet:2002gs} (see Eqs.~\eqref{regulators}).
	\item For each regulator family and at a given order of the DE, one varies the value of $\alpha$ and looks for an extrema of the quantity $Q$ ($\alpha_{PMS}$).
	\item When comparing among different families of regulating functions we choose as estimate at order $\mathcal{O}(\partial^{s})$:
	\begin{equation}\label{eq:regEst}
		\bar{Q}^{(s)}=\frac{{max}_i\lbrace Q_{Reg;i}^{(s)}\rbrace+{min}_i\lbrace Q_{Reg;i}^{(s)}\rbrace}{2}.
	\end{equation}
	\item Having determined the $\bar Q^{(s)}$ at various orders, we consider as the error estimate of the DE at order $\mathcal{O}(\partial^s)$, $\bar\Delta Q^{(s)}=|\bar Q^{(s)}-\bar Q^{(s-2)}|/4$.  The 1/4 corresponds to the more conservative estimation of the small parameter.
\end{itemize}
This procedure was tested successfully for critical exponents in Ref.~\cite{DePolsi:2020pjk}. The test is successful in two ways. First, the predictions so formulated are {\it accurate}: whenever the most precise prediction is not the one obtained from the DE, it is within error bars of the DE. Second, it is {\it self-consistent} in the sense that predictions made at order $\mathcal{O}(\partial^{4})$ are always within error bars of predictions made at the previous order $\mathcal{O}(\partial^{2})$.

Moreover, it has been observed that, in many cases, those predictions are somewhat pessimistic. In particular, when the results seem to oscillate at successive orders around the correct value, one can improve significantly the estimate of central values and error estimates. Unfortunately, the UAR do not seem to show any clear oscillatory behavior and, consequently, we limit here to this relatively conservative estimate of error bars.

The previous estimates concern the systematic error coming from truncating the DE at a finite order. It is necessary to take into account also other independent sources of error. First, there is the dependence of the results among the various families of regulators. This source of error is typically smaller than the one previously mentioned but since it is not negligible we must take it into consideration. The error associated with the dispersion in regulators $\Delta Q_{reg}$ is taken to be the width of this dispersion. A second source of error comes from the fact that, as we show in Section~\ref{results}, the results for the high temperature UAR are computed from a plateau. However, for some UAR this plateau presents a non-negligible variation which we take into account as another source of error $\Delta Q_{num}$ (this variation is typically the same for different regulators, so there's no need to make any precision on this point). Moreover, even if these estimates are typically pessimistic they can become too optimistic in the exceptional case where two consecutive orders of the DE accidentally cross. This has been analyzed in detail in Ref.~\cite{DePolsi:2020pjk} and we refer the reader to this reference for a detailed discussion on this point. In the present work we have one such quantity denoted in the literature as $r_8$. We discuss the specific difficulties in calculating its error bars in Section~\ref{results}.

To summarize, we consider the estimate of a quantity with the DE at order $\mathcal{O}(\partial^l)$ to be $$\bar{Q}^{(l)}\pm \Delta \bar{Q}^{(l)},$$ with  $\bar{Q}^{(l)}$ given by Eq.\eqref{eq:regEst} and the error is computed as $$\Delta \bar{Q}^{(l)}=\Delta Q_{reg}+\Delta Q_{num}+\bar{\Delta} Q^{(l)},$$ with $\Delta Q_{num}$ the variation of the quantity along the plateau, $$\Delta Q_{reg}={max}_i\lbrace Q_{Reg;i}^{(l)}\rbrace-{min}_i\lbrace Q_{Reg;i}^{(l)}\rbrace,$$ and $$\bar\Delta Q^{(s)}=|\bar{Q}^{(s)}-\bar{Q}^{(s-2)}|/4.$$

\subsection{Calculation of UARs within the NPRG}

The goal of the work done in this article is to calculate the UARs we have described in the previous section by numerically solving the NPRG  Eq.~\eqref{wettericheq} for the effective action $\Gamma_k$. As mentioned previously this is done by taking the derivative expansion ansatz for $\Gamma_k$ described in equation \eqref{ansatz-order4}, which is determined by a set of functions $U_k(\rho)$, $Z_k(\rho)$, $Y_k(\rho)$ and $W_I(\rho)_{I=1,\dots,10}$. The RG flow for these functions is given by the NPRG equations, and, as described in more detail in Section \ref{results}, is solved numerically by discretizing the variable $\rho$ and working with a large but finite set of ODEs. In order to calculate the universal quantities described in this section, the RG flow needs to be solved for initial conditions of the form of Eq.~\eqref{linearflowui}; that is, tuned so that the RG flow passes sufficiently close to the trajectories joining the WF fixed point to either the high temperature or low temperature fixed points. Once the RG flow of $\Gamma_k$ is obtained for these initial conditions one can extract the leading scaling behavior in $t$ for quantities such as the susceptibilities $\chi_{2n}$ from the ``long RG-time'' behavior of $\Gamma_k$. From here one can, in principle, calculate the UARs we have described for $O(N)$ models within the DE approximation scheme. In fact, in order to do this it is possible to express the UARs directly in terms of the functions characterizing the DE ansatz Eq.~\eqref{ansatz-order4}. For instance, we have
\begin{equation}\label{g4fromDE}
g_4 = -3\frac{U''(0)}{(U'(0))^{2-d/2}Z(0)^{d/2}}.
\end{equation}
The derivation of this expression for $g_4$, as well as analogous formulas for other UARs, is straightforward by substituting the ansatz for the effective action in Eq.\eqref{eq:g_4FromFs}. Please note, however, that the expression in equation \eqref{g4fromDE} is independent of the order of the DE (even if the actual values of the functions \textit{does} depend on the order of the approximation). Analogous formulas for other UARs can be found in Appendix \ref{appendixhifhtUar}.\\ 

\subsection{Approximations for the Equation of State}

From formulas such as Eq.~\eqref{g4fromDE} we see that calculation of UARs would appear to be relatively straightforward once the ``long RG-time'' behavior of the $U_k(\rho)$, $Z_k(\rho)$, $Y_k(\rho)$ and $W_I(\rho)_{I=1,\dots,10}$ is obtained. However, it turns out that due to technical difficulties in the numerical solution of the RG flow equations this is not the path that we will take here to compute many of the UARs we have described in this section. As we have seen, many of the universal quantities we are interested in involve a ratio of amplitudes calculated in different phases, so that the ``long RG-time'' behavior of $\Gamma$ would need to be obtained both at $t>0$ as well as at $t<0$. It is well known in the NPRG literature however \cite{Berges:2000ew,Pelaez:2015nsa,Dupuis:2020fhh} that the flow in the low temperature region passes close to a point where the propagator diverges and the equations become singular, which complicates the numerical integration of the flow. This singularity is related to the approach of a convex non-analytical effective potential in the low temperature phase. Furthermore, in this phase in order to extract from $\Gamma_k$ the scaling behavior of the quantities involved it is necessary to use dimensionful variables instead of the standard dimensionless ones when integrating the flow, once the flow approaches the low temperature fixed point. This significantly adds to the computational complexity and cost of the numerical integration.

In order to avoid these difficulties we have opted to take an alternate route which we presently describe. As explained in more detail in subsequent sections, we only solve the flow in the high temperature region. From this, we obtain the long RG-time behavior of the functions $U_k(\rho)$, $Z_k(\rho)$, $Y_k(\rho)$ and $W_I(\rho)_{I=1,\dots,10}$ characterizing $\Gamma_k$ for trajectories that pass very close to the RG orbit joining the WF fixed point with the high-temperature fixed point. From here, through the method discussed in previous paragraphs we may extract from the long RG-time behavior of $U_k(\rho)$ and $Z_k(\rho)$ the values of the fully high-temperature UARs $g_4$, $r_6$, $r_8$ and $r_{10}$. These UARs furnish an approximate representation of the function $G(z)$ defined in Eq.~\eqref{highTEOS} and thus give an approximation to the universal equation of state of the system in the high temperature regime.

In order to calculate the other UARs, which involve quantities in the low-temperature, critical and crossover regions, which we do not probe numerically, we employ a method that is frequently used in the literature. This involves finding a way to continue our approximate representation of the EOS of the system through the critical temperature and into the low temperature region. From such an (approximate) expression for the EOS in those regimes it is then possible to extract the remaining UARs.

Let us describe this procedure in more detail. As has been mentioned, this method has been frequently used and can be found in several references \cite{Pelissetto02,ZinnJustin:2002ru}. We reproduce it here to make the presentation self-contained. The key point is to find an approximate representation of the EOS that holds in the low temperature region. Notice that Eq.~\eqref{highTEOS} is only valid in the high temperature regime. However, using scaling arguments similar to those we have explained earlier, it turns out that the EOS can be written also in the form
\begin{equation}
H_i = M_i B_c^{-\delta} M^{\delta-1}f(x),
\end{equation}
for some universal function $f(x)$ where
\begin{equation}
x = B^{1/\beta}t M^{-1/\beta}.
\end{equation} 
Crucially, this expression for the equation of state holds both above and below $t=0$, and the remaining UARs that we are considering may be obtained in terms of it. Intuitively,  the procedure we will follow consists in using our approximate expression for $G(z)$ to find an approximate expression for $f$ in the region where they overlap, and then to continue this expression for $f$ into the low temperature region. To accomplish this while at the same time imposing that the analytic structure of $f$ is properly taken into account, the standard way to do this is to use a two-parametric representation of the variables $H$, $t$ and $M$. One writes these in terms of the two variables $\theta$ and $R$ as follows
\begin{align}
H = h_0R^{\beta\delta}h(\theta),\\
t = R(1-\theta^2),\\
M = m_0R^{\beta}m(\theta),
\end{align}
where $R$ is restricted to take positive values only and $h_0$ and $m_0$ are factors related to the normalization of $h(\theta)$ and $m(\theta)$. The functions $h(\theta)$ and $m(\theta)$ must be odd in order to guarantee that only odd powers of $M$ appear in the high temperature EOS, i.e. that $G(z)$ defined by Eq.~\eqref{highTEOS} is an odd function of $z$. The different regimes of interest for the study of the thermodynamics of the system appear in this parametrization in the following way. Since $h(\theta)$ is an odd function, the $\theta=0$ curve describes the high temperature phase of the system at zero external field. The $t<0$ phase at zero external field corresponds to the curve at $\theta = \theta_0$, $\theta_0$ being the lowest positive root of $h(\theta)$, which must be larger than $1$ in order for this to indeed occur at negative reduced temperature. Finally, the critical isotherm corresponds to the curve $\theta=1$. Notice that the dependence on $R$ in the parametrization is chosen so that the scaling relations involving the magnetization in these last two regimes are automatically satisfied for arbitrary $m(\theta)$ and $h(\theta)$. Of course, the functions $m(\theta)$ and $h(\theta)$ are not arbitrary bur rather they must be chosen in such a way that the equation of state is satisfied in the full $(t,H)$ plane. This is accomplished by imposing that
\begin{equation}\label{foftheta}
f(x(\theta)) =  \left(\frac{m(\theta)}{m(1)}\right)^{-\delta}\frac{h(\theta)}{h(1)},
\end{equation}
where $x$ and $\theta$ are related by
\begin{equation}\label{xoftheta}
x(\theta) = \frac{1-\theta}{\theta_0^2-1}\left(\frac{m(\theta_0)}{m(\theta)}\right)^{1/\beta}.
\end{equation}
This requirement can also be formulated for $t>0$ in terms of $G(z)$, which yields
\begin{equation}\label{Goftheta}
G(z(\theta)) =  \sigma\, h(\theta)(1-\theta^2)^{-\delta\beta},
\end{equation}
where $z$ and $\theta$ are related by
\begin{equation}\label{zoftheta}
z(\theta) = \sigma\, m(\theta)(1-\theta^2)^{-\beta},
\end{equation}
where $\sigma$ is a free parameter related to the overall normalization of the functions determining the parametrization. Notice that this last equation only yields a real $z$ in the $\theta<1$ region, which corresponds to the high temperature phase.

The usefulness of this representation of the equation of state is that the functions $m(\theta)$ and $h(\theta)$ are analytic in the physical range $0\leq\theta<\theta_0$, due to the fact that the equation of state is expected to be analytic everywhere except at the critical point $(t,H)=(0,0)$ (which corresponds to $R=0$ in this parametrization) and at the coexistence curve $(t<0,H=0)$, which corresponds to $\theta = \theta_0$. This suggests that a possible approximation scheme to extract the EOS from our knowledge of the high temperature physics will be to take $m(\theta)$ and $h(\theta)$ analytic functions. Since we calculate only a finite number of the coefficients $r_i$ of the polynomial expansion of $G(z)$ at $z=0$, we have only knowledge of the function $G(z)$ at small $z$ (that is, in the small $H$ regime at $t>0$). In the $(R,\theta)$ parametrization this corresponds to the small $\theta$ region. In practice, when trying to determine $m(\theta)$ and $h(\theta)$ we can therefore only impose that they satisfy Eq.~\eqref{Goftheta} and Eq.~\eqref{zoftheta} in this region, which amounts to imposing conditions on a finite number of terms in the polynomial expansions of these functions around $\theta=0$. The approximation we will use then consists of taking the finite polynomial obtained from solving these conditions to be a good description of these functions --and therefore of the EOS-- in the whole $\theta$ range.

We then employ the following strategy. One calculates an approximate representation of $G(z)$ in terms of the $r_i$ we have calculated by truncating the polynomial expansion. One then takes a polynomial ansatz for $h(\theta)$ and $m(\theta)$ depending on some free parameters, of equal number as the number of $r_i$ used to approximate $G(z)$. Then the free parameters of our ansatz for $m(\theta)$ and $h(\theta)$ are obtained by imposing that Eq.~\eqref{Goftheta} and Eq.~\eqref{zoftheta} are satisfied. In fact, doing this order by order in an expansion around $\theta=0$ yields a series of algebraic equations that completely determine our free parameters and may be easily solved. This procedure determines the parametrization. With this, one reconstructs $f(x)$ and from here the remaining UARs are calculated. As we have mentioned before, this procedure is by now standard and so there exist in the literature formulas which allow to calculate the UARs directly from the functions $m(\theta)$ and $h(\theta)$ defining our parametrization. In Appendix \ref{appParam} we reproduce these and provide a more detailed exposition of the procedure just described.

Before we proceed to describe the results of our calculations, a few remarks are in order. First, although the use of the parametrization we have described as well as its polynomial approximation is frequent in the literature, the methods employed to determine the coefficients involved vary. The approach we have decided to follow here of choosing the coefficients so as to reproduce the low-$z$ behavior of $G(z)$ is essentially that of \cite{Campostrini_2002}. For other approaches we refer to \cite{Pelissetto02} and references therein. 

Second, as can be appreciated in the preceding formulae, the method employed requires as an inputs the values of the critical exponents, which are not calculated in this work. For consistency, we have chosen to use the values of these as obtained from the NPRG calculations performed in \cite{DePolsi:2020pjk}. Since we calculate the high temperature UARs at different orders in the DE and take the different results as inputs to the parametrizat1ion method, we have been careful to input the critical exponents calculated at the same order in the DE, e.g. when using the $r$s and $g_4$ obtained at second order in the DE to calculate the other UARs we use the critical exponents at second order from \cite{DePolsi:2020pjk} in this calculation.

Third, we will only use the parametrization approach previously described to calculate UARs for $N=\{2,3,4,5\}$. For $N=1$ the parametrization is somewhat different and will be omitted for simplicity. In any case, in regards to the work presented here, the point of calculating the UARs that are not exclusively high-temperature is mainly to establish whether they can be obtained from the NPRG calculation in the high-temperature regime through the approximate parametrization procedure or whether a NPRG study of the low-temperature phase is warranted. This last point will be discussed further in the conclusions.

\section{Results}\label{results}

In this section we use as an example the $O(2)$ model and present first the typical curves for $g_4$ and the $r_{n}$'s obtained with a given regulator as a function of the RG time $s$ along with the procedure we use to obtain from this the data of interest and an associated error to it. After this we present the curves of these high-temperature UARs as a function of the regulator parameter $\alpha$ which resemble the ones presented in Ref.~\cite{DePolsi:2020pjk} for the critical exponents. As described in previous sections, all the UARs we compute can be obtained from $g_4$ and the $r_{n}$ in conjuntion with critical exponents. We then present the results obtained for all considerd UARs for the studied $O(N)$ models. Finally, we present the results for the UARs involving low temperature amplitudes.

\subsection{Procedure and associated errors to UARs}

We start by stressing that the general procedure for obtaining or approaching the fixed point is the same as the one described in Ref.~\cite{DePolsi:2020pjk}. However, from this point forward, we must follow the flow along or close to the heteroclinic line that conects the critical fixed point to the high temperature fixed point. At some point along this flow, the numerical error starts to grow due to the sharpness of the potential around the stable minimum at zero magnetization.\footnote{This is due to the fact that we are using dimensionless quantities for the computations.} Nonetheless, this ocurs late in the flow and quantities stabilize just enough for their computation. Since the $r_{n}$ are computed through derivatives of the potential at this particular point (see Eqs.~\eqref{r6eq}-\eqref{r10eq}), the error associated to these quantities grow with $n$.

\begin{figure}[!ht]
	\begin{center}
		\includegraphics[width=0.45\textwidth]{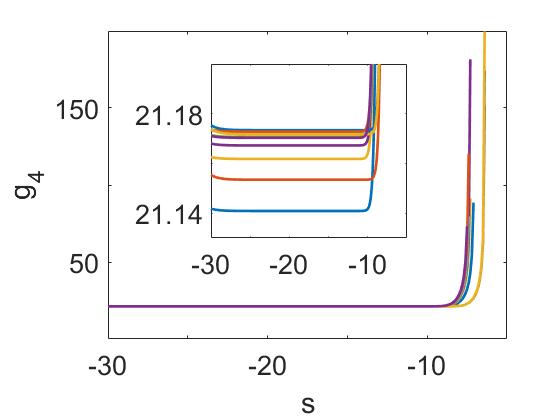}
	\end{center}
	\caption{Dependence of UAR $g_4$ as a function of the RG time $s$ with regulator $E_k$ for the $O(2)$ at order $\mathcal{O}(\partial^4)$ of the DE. Different colors correspond to different $\alpha$ values evenly spaced in the interval $[1,2]$.}
	\label{figG4}
\end{figure}

\begin{figure}[!ht]
	\begin{center}
		\includegraphics[width=0.45\textwidth]{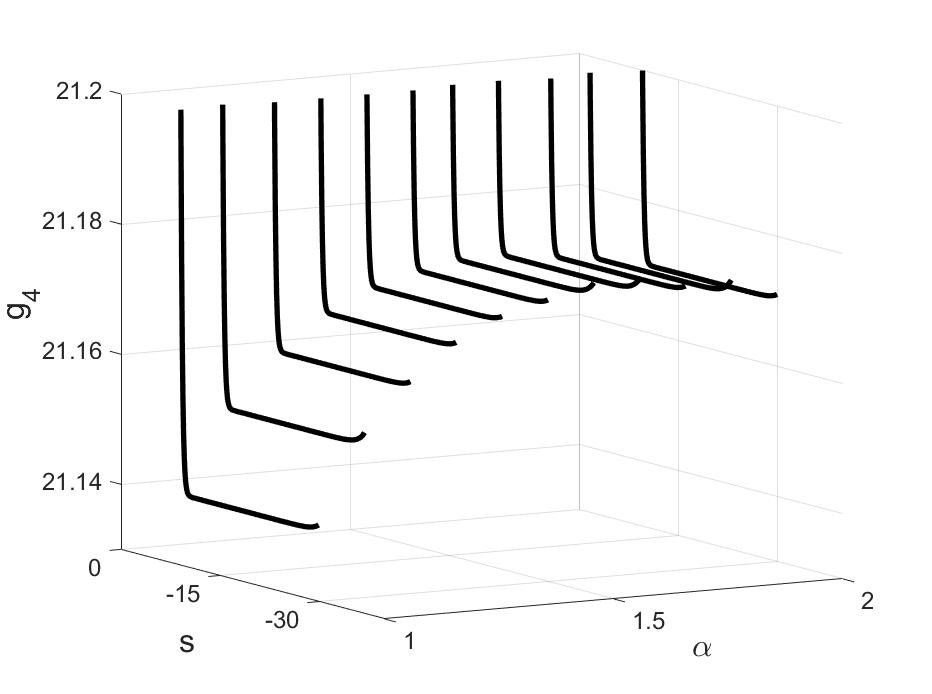}
	\end{center}
	\caption{Exhibiting the PMS behaviour for the UAR $g_4$ as a function of the RG time $s$ and $\alpha$ parameter with regulator $E_k$ for the $O(2)$ at order $\mathcal{O}(\partial^4)$ of the DE.}
	\label{figG4b}
\end{figure}

\begin{figure}[!ht]
	\begin{center}
		\includegraphics[width=0.45\textwidth]{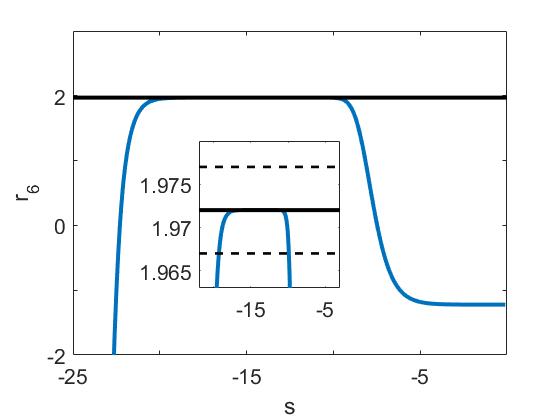}
	\end{center}
	\caption{The UAR $r_6$ as a function of the RG time $s$ corresponding to the PMS value of $\alpha$ with regulator $E_k$ for the $O(2)$ model at order $\mathcal{O}(\partial^4)$ of the DE (blue solid line). Estimate of this quantity with the DE at order $\mathcal{O}(\partial^4)$ (black solid line) and error bars (black dashed line) are also introduced for reference. A zoom on the region of interest is also introduced for visualisation purposes.}
	\label{figR6}
\end{figure}

\begin{figure}[!ht]
	\begin{center}
		\includegraphics[width=0.45\textwidth]{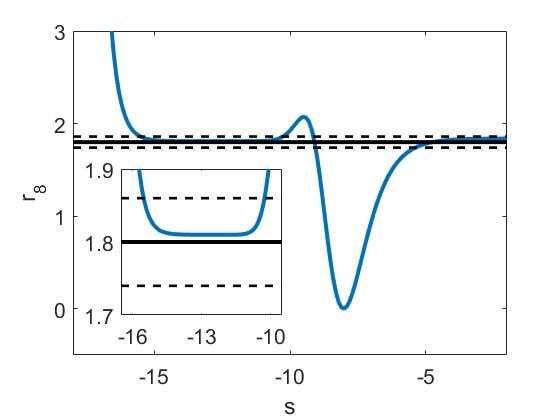}
	\end{center}
	\caption{The UAR $r_8$ as a function of the RG time $s$ corresponding to the PMS value of $\alpha$ with regulator $E_k$ for the $O(2)$ model at order $\mathcal{O}(\partial^4)$ of the DE (blue solid line). Estimate of this quantity with the DE at order $\mathcal{O}(\partial^4)$ (black solid line) and error bars (black dashed line) are also introduced for reference. A zoom on the region of interest is also introduced for visualisation purposes.}
	\label{figR8}
\end{figure}

\begin{figure}[!ht]
	\begin{center}
		\includegraphics[width=0.45\textwidth]{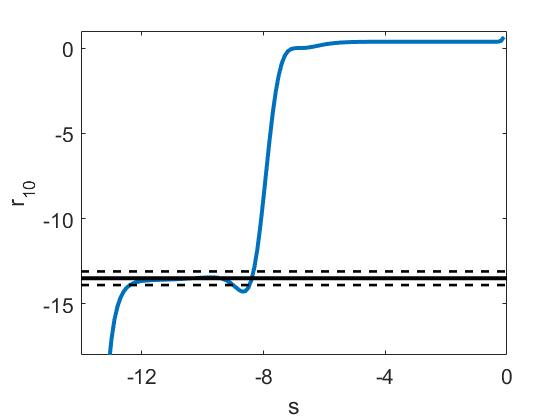}
	\end{center}
	\caption{The UAR $r_{10}$ as a function of the RG time $s$ corresponding to the PMS value of $\alpha$ with regulator $E_k$ for the $O(2)$ model at order $\mathcal{O}(\partial^4)$ of the DE (blue solid line). Estimate of this quantity with the DE at order $\mathcal{O}(\partial^4)$ (black solid line) and error bars (black dashed line) are also introduced for reference.}
	\label{figR10}
\end{figure}

We start by presenting $g_4$ as a function of the RG time $s$ with the regulator given in Eq.~\eqref{regulator-wetterich} for the $O(2)$ model, which is representative of other regulators and other $O(N)$ models as well. As we flow away from the critical fixed point towards the high temperature fixed point, the curve for $g_4$ behaves similarly for different orders of the DE and we focus on the curves corresponding to order $\mathcal{O}(\partial^4)$. This is shown in Fig.~\ref{figG4} where different curves correspond to different $\alpha$ values. In this figure we can see how $g_4$ converges consistently to a plateau near the expected values. Additionally, we show in Fig.~\ref{figG4b} the smooth dependence of the height of the plateau with the regulator parameter $\alpha$.
It is worth emphasizing that the origin of the RG time $s$ is taken to be $0$ not at the microscopic scale but rather already around the critical fixed point. Otherwise one could need even larger RG times to reach this plateau. Nevertheless, considering these perturbations as the microscopic theories, the RG time $s\sim-8$ corresponds to approximately the inverse of the correlation length $\xi$. It is already evident that around $s\sim-30$ the numerical error is starting to affect the computation of $g_4$. The situation enhances as we consider the UARs $r_{n}$ with increasing $n$. We show this phenomena through Figs.~\ref{figR6}--\ref{figR10} corresponding to $r_6$, $r_8$ and $r_{10}$, respectively. It should be noticed that there is another plateau for $s\gtrsim-5$ which corresponds to our choice of initial conditions that corresponds to a system very close to the WF fixed point. This makes the flow very slow in the initial stages of the RG evolution. The extension of the plateau of interest, for $s\lesssim-8$, as well as its quality manifestly deteriorates. Although one could increase the stability of these plateaus by implementing a higher control over the flows, this is of no benefit since, as it will become evident, the error associated to the finite order of the DE is generally dominant.


When the plateau is well behaved it is easy to assign a value since variation along many units of RG time is negligible in comparison with other sources of error. However, to assign a value to the quantity when the plateau has a slight tilt or a wavy behavior on top of it, we choose the situation with least slope well inside the plateau (this is because the plateau may begin or end with a small maximum or minimum as can be seen in Fig.~\ref{figR10}). This contributes as an extra source of error which we consider to be the variation of the quantity along the plateau. This error is not dominant but in certain cases it is non-negligible. In particular, it must be taken into consideration for the computation of $r_{10}$. 

\begin{figure}[!ht]
	\begin{center}
		\includegraphics[width=0.45\textwidth]{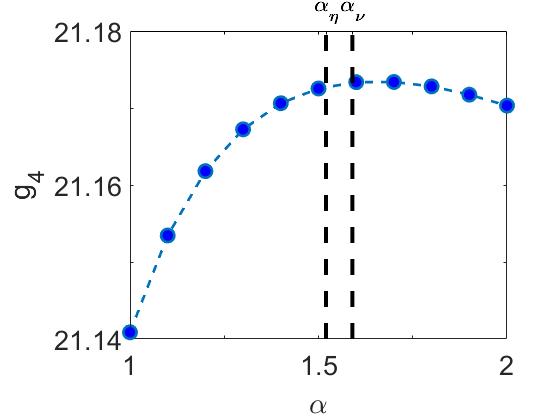}
	\end{center}
	\caption{$g_4$ vs $\alpha$ for $O(2)$ with the regulator $W_k$.}
	\label{figG4vsalp}
\end{figure}

\begin{figure}[!ht]
	\begin{center}
		\includegraphics[width=0.45\textwidth]{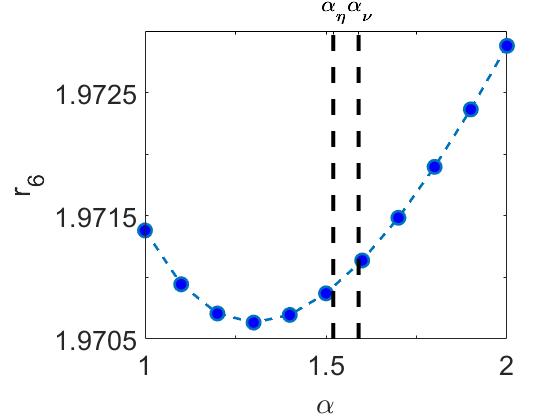}
	\end{center}
	\caption{$r_6$ vs $\alpha$ for $O(2)$ with the regulator $W_k$.}
	\label{figR6vsalp}
\end{figure}

\begin{figure}[!ht]
	\begin{center}
		\includegraphics[width=0.45\textwidth]{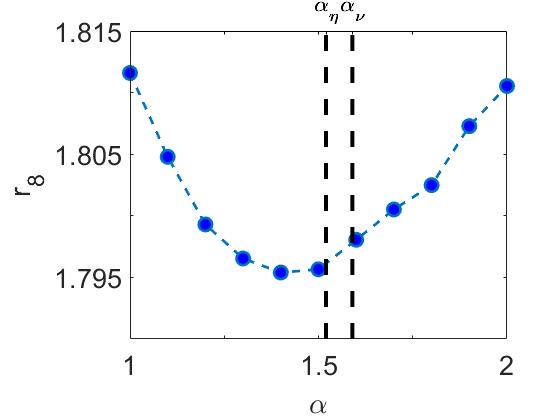}
	\end{center}
	\caption{$r_8$ vs $\alpha$ for $O(2)$ with the regulator $W_k$.}
	\label{figR8vsalp}
\end{figure}

\begin{figure}[!ht]
	\begin{center}
		\includegraphics[width=0.45\textwidth]{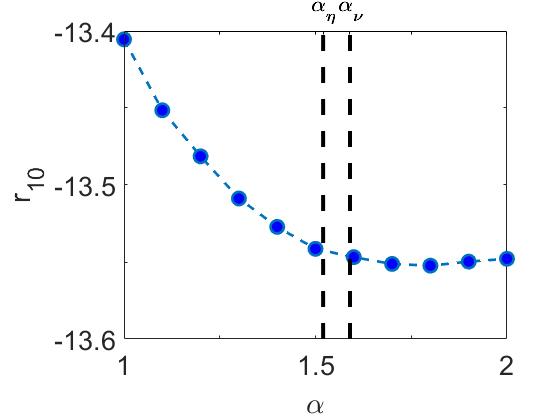}
	\end{center}
	\caption{$r_{10}$ vs $\alpha$ for $O(2)$ with the regulator $W_k$.}
	\label{figR10vsalp}
\end{figure}

The procedure from this point onward is completely equivalent to what is done for the critical exponents in \cite{DePolsi:2020pjk} as far as concerns assigning definite values to quantities. For completeness we show in Figs.~\ref{figG4vsalp}--\ref{figR10vsalp} the quantities of interest when varying the regulator scale parameter $\alpha$.

The behavior of all these quantities is very similar to those exhibited by the critical exponents. In particular, the values of $\alpha$ at which the different quantities present a maximum or a minimum according to the PMS criterion, or a behaviour pursuing the PMS spirit,\footnote{We refer the reader to Ref.~\cite{DePolsi:2020pjk} for a detailed discussion of these quantities.} are very close to the values of $\alpha_{PMS}$ at which the critical exponents $\eta$, $\nu$ and $\omega$ present theirs. Indeed, the values of $\alpha_{PMS}$ for $\eta$, $\nu$ and $\omega$ for the $O(2)$ universality class at order $\mathcal{O}(\partial^4)$ with the regulator $W_k$ are $\alpha_{\eta}=1.52(1)$, ${\alpha_{\nu}=1.59(1)}$ and $\alpha_{\omega}=1.59(1)$, respectively. These values are to be compared with the values of $\alpha$ at which these high-temperature UARs present their extrema. We have included for reference in Figs.~{\ref{figG4vsalp}-\ref{figR10vsalp}}, the values of $\alpha_{\eta}$ and $\alpha_{\nu}$. After a quick examination of these figures it becomes evident that the optimal value of $\alpha$ (according to the PMS criterion) for all the high-temperature UARs studied in this work lay around $1.5\pm0.2$.

\begin{figure}[!ht]
	\begin{center}
		\includegraphics[width=0.45\textwidth]{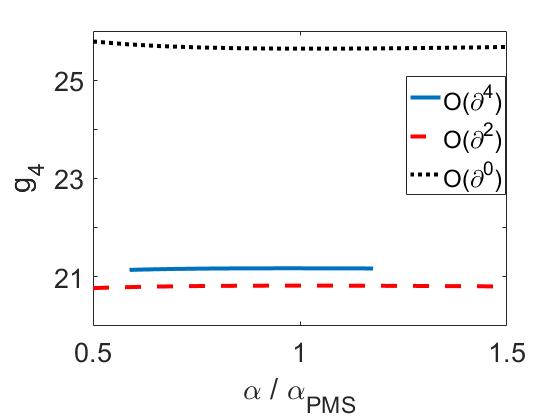}
	\end{center}
	\caption{$g_4$ vs $\alpha/\alpha_{PMS}$ for $O(2)$ with the regulator $W_k$.}
	\label{g4vsalpnormDE}
\end{figure}

Moreover, we show that these quantities also behave similarly to critical exponents as we change the order of the DE as expected. We exemplify this using the UAR $g_4$ and showing its dependence on the $\alpha$ parameter in Fig.~\ref{g4vsalpnormDE}. It is evident that as we consider higher orders of the DE results seems to converge yielding reasonable results.

Finally, from the PMS values at every considered order and for each regulator we compute our estimated quantities as explained in Sec.~\ref{errorbars}. The tables with the UAR thus obtained for the regulators considered, at the different orders of the DE, can be found in Appendix~\ref{Ap:NumParam}.

\subsection{High-temperature UARs for $N=1,2,3,4$ and $5$}

We now present the results for the many universal amplitude ratios that follow from the procedure described previously. We start by recalling this procedure and the details of the calculations. We have considered some of the regulators employed previously in \cite{DePolsi:2020pjk} with the same truncation of the DE. The specifics of the numeric integration of the flow can be found in Appendix~\ref{App:Num}). 

At this point, we would like to make an important remark. The general method proposed for assigning error bars with the DE to the different quantities lays on the fact that the difference of quantities at successive orders of the DE do not become accidentally small. This situation already occurs for the $\omega$ exponent and was overcame by imposing a monotonically decreasing behavior with $N$ as we take $N\to\infty$. This same behavior is encountered for the $r_8$ UAR when comparing orders $\mathcal{O}(\partial^0)$ and $\mathcal{O}(\partial^2)$. However, there are two reasons why we can not implement the same program. On the one hand, we did not considered sufficiently high values of $N$ so as for this to be a reasonable criterion. On the other hand, and more importantly, we did take a glance at the behavior at higher $N$ and the situation is that this anomalous behavior continues to be present for large values of $N$. This implies that by the time the difference between orders $\mathcal{O}(\partial^0)$ and $\mathcal{O}(\partial^2)$ becomes sufficiently large in comparison with the difference of orders $\mathcal{O}(\partial^2)$ and $\mathcal{O}(\partial^4)$, the large $N$ suppression of errors already kicked in and we cannot fix the accidentally small error bars at small values of $N$ for the $r_8$ UAR.  This only means that there exists quantities for which the proposed method for assigning error bars with the DE is not valid at certain orders of the DE where accidental crossing at successive orders exists. 

The results are presented and compared to the best results of the literature. We refer the reader to \cite{Pelissetto02} for an extensive reference on the documented results.

We consider the universality classes of models $O(N)$ with $N\in \lbrace 1,2,3,4,5\rbrace$, which we briefly describe below.

We start by discussing the results for the $\mathbb{Z}_2$ or $O(1)$ universality class which corresponds to the Ising model, pure substances in their liquid-gas transition or uniaxial magnets to name a few systems. For this particular universality class there is a plethora of results but, as we have already stated, we will only present the most precise ones.

\begin{table*}
	\caption{\label{T:O1}High temperature UARs for $N=1$. The error bar for $r_8$ marked with an asterisk is under-estimated, see main text.}
	\begin{ruledtabular}
		\begin{tabular}{lllll|c c c}
			\multicolumn{1}{c}{}& 
			\multicolumn{1}{c}{HT}& 
			\multicolumn{1}{c}{$\epsilon$ exp}& 
			\multicolumn{1}{c}{$d=3$ exp} &
			\multicolumn{1}{c}{MC} &
			\multicolumn{1}{c}{LPA}&
			\multicolumn{1}{c}{$\mathcal{O}(\partial^2)$}&
			\multicolumn{1}{c}{$\mathcal{O}(\partial^4)$}\\
			\hline  
			
			$g_4$ &23.56(2) \cite{Campostrini:2002cf}&  23.6(2) \cite{pelissetto2000effective}   & 23.64(7) \cite{Guida:1998bx}&23.6(2) \cite{kim2000critical}& 29.2 & 23.1(16) & 23.60(15)\\
			& &  &&23.4(2) \cite{pelissetto1998effective,ballesteros1998finite}&21(4) \cite{morris1997three}&&\\
			
			$r_6$ &  2.056(5)  \cite{Campostrini:2002cf} & 2.058(11) \cite{pelissetto1998effective} & 2.053(8) \cite{Guida:1998bx} 
			& 2.72(23) \cite{tsypin1994universal} & 2.0 & 2.05(1) & 2.064(6)\\
			&   1.99(6) \cite{butera19972n} & 2.12(12) \cite{Guida:1998bx} & 2.060
			\cite{sokolov1999universal} & 3.37(11) \cite{kim2000critical} & 2.064(36) \cite{morris1997three}&&\\
			&   2.157(18) \cite{zinn1996renormalized} &                         &                 &         3.26(26)~\cite{kim1997monte}  &&&\\
			&   2.25(9) \cite{nan1996criticality} & & & &  &&  \\
			
			$r_8$ & 2.3(1) \cite{Campostrini:2002cf}  & 2.48(28) \cite{pelissetto1998effective} & 2.47(25) \cite{Guida:1998bx} 
			& & 2.64 & 2.40(6*) & 2.60(4)\\
			&  2.7(4) \cite{butera19972n} & 2.42(30) \cite{Guida:1998bx} &  & & 2.47(5) \cite{morris1997three}&&\\
			
			$r_{10}$ & 
			$-$13(4) \cite{campostrini1999improved}& $-$20(15) \cite{pelissetto1998effective} & $-$25(18) \cite{Guida:1998bx} 
			&  & -9.5 & -14.8(14) & -14.1(3)\\
			&  $-$4(2) \cite{butera19972n} &  $-$12.0(1.1) \cite{Guida:1998bx} & &   & $-$18(4) \cite{morris1997three}&&\\
		\end{tabular}
	\end{ruledtabular}
\end{table*}

As can be seen from Table \ref{T:O1}, the results obtained for the different UARs with the DE are compatible with most of the results reported in the literature. Moreover, the attained precision is an order of magnitude worse for the UAR $g_4$, but is similar or better for the remaining UARs. 
It is interesting to observe that our result for $r_6$  is compatible with other field-theoretical results but that all Monte-Carlo results for this quantity are incompatible with these methods (that claim smaller error bars).

These results show that the method is very versatile and accurate and can be implemented to compute many quantities of interest.

The $O(2)$ model corresponds to the XY model, easy plane magnets or the Helium-4 in its fluid-superfluid $\lambda$ transition. The results obtained for the $O(2)$ universality class, presented in Table \ref{T:O2} show that the precision obtained with this method for the $g_4$ UAR is of the same order of magnitude but a bit less precise than other results found in the literature. However, in contrast with the $\mathbb{Z}_2$ universality class, the precision obtained for the $O(2)$ universality class for the remaining UAR is always higher than with any other method.

\begin{table*}
	\caption{\label{T:O2}High temperature UARs for $N=2$. The error bar for $r_8$ marked with an asterisk is under-estimated, see main text.}
	\begin{ruledtabular}
		\begin{tabular}{clll|ccc}
			\multicolumn{1}{c}{}& 
			\multicolumn{1}{c}{HT}& 
			\multicolumn{1}{c}{$d=3$ exp}&
			\multicolumn{1}{c}{$\epsilon$ exp} &
			\multicolumn{1}{c}{LPA} &
			\multicolumn{1}{c}{$\mathcal{O}(\partial^2)$} &
			\multicolumn{1}{c}{$\mathcal{O}(\partial^4)$} \\
			\hline  
			$g_4^+$   & 21.14(6) \cite{campostrini2001critical} & 21.16(5)  \cite{Guida:1998bx} & 
			21.5(4) \cite{pelissetto2000effective,pelissetto1999non}& & &\\
			& 21.05(6) \cite{campostrini2000critical} &  21.20(6) \cite{le1980critical} && 25.7 & 20.8(12) & 21.18(10) \\
			&    & &&&& \\
			&   & & &&&\\
			
			$r_6$ & 1.950(15) \cite{campostrini2001critical}  &   1.967 \cite{sokolov1999universal} & 1.969(12) \cite{pelissetto2000effective,pelissetto1998effective} & 1.91 & 1.96(1) & 1.972(5)\\
			& 1.951(14) \cite{campostrini2000critical}& &   &&&\\
			&  & &   &&&\\
			
			$r_8$       & 1.44(10) \cite{campostrini2001critical}   &  1.641 \cite{sokolov1999universal} & 2.1(9) \cite{pelissetto2000effective,pelissetto1998effective} & 1.79 & 1.64(4*) & 1.80(6)\\
			& 1.36(9)\cite{campostrini2000critical}      & &  &&&\\ 
			
			$r_{10}$ & $-$13(7) \cite{campostrini2001critical}   & &  & -9.47 & -14.2(15) & -13.5(4)\\
		\end{tabular}
	\end{ruledtabular}
\end{table*}

We now present the results for the remaining studied universality classes which are the $O(3)$, $O(4)$ and $O(5)$ models. The $O(3)$ is commonly related to the universality class of isotropic magnets and the $O(4)$ model is associated to the chiral transition of quantum chromodynamics with two flavors.

\begin{table*}
	\caption{\label{T:O3}high temperature UARs for $N=3$. The error bar for $r_8$ marked with an asterisk is under-estimated, see main text.}
	\begin{ruledtabular}
		\begin{tabular}{clll|ccc}
			\multicolumn{1}{c}{}& 
			\multicolumn{1}{c}{HT}& 
			\multicolumn{1}{c}{$d=3$ exp}&
			\multicolumn{1}{c}{$\epsilon$ exp} &
			LPA&
			$\mathcal{O}(\partial^2)$&
			$\mathcal{O}(\partial^4)$\\
			\hline  
			
			$g_4^+$   & 19.13(10) \cite{Campostrini_2002} & 19.06(5) \cite{Guida:1998bx} &
			19.55(12) \cite{pelissetto2000effective,pelissetto1999non}  & 22.6 & 18.9(9) & 19.10(6)\\
			& 19.31(14),$\;$ 19.27(11) \cite{butera1998renormalized}  &  19.06 \cite{murray1991revised} & &22.35 \cite{Berges:1995mw,Berges:2000ew}&&\\ 
			& 19.34(16) \cite{pelissetto1999non}  & &  &  &&\\ 
			
			$r_6$   & 1.86(4) \cite{Campostrini_2002}  & 1.880 \cite{sokolov1999universal} &
			1.867(9) \cite{pelissetto2000effective,pelissetto1998effective} & 1.8 & 1.88(2) & 1.886(3)\\
			& 2.1(6) \cite{reisz1995high} & 1.884(32) \cite{pelissetto2000effective} & &  1.74 \cite{Tetradis94} &&\\ 
			
			$r_8$   & 0.6(2)\cite{Campostrini_2002}  & 0.975 \cite{sokolov1999universal} &
			1.0(6) \cite{pelissetto2000effective,pelissetto1998effective}& 1.063 & 1.02(1*) & 1.14(4)\\
			&&& & 0.84 \cite{Tetradis94}&&\\
			
			$r_{10}$   & $-$6(3)\cite{Campostrini_2002} & &  & -8.05 & -12.5(14) & -12.1(3)\\
		\end{tabular}
	\end{ruledtabular}
\end{table*}

\begin{table}
	\caption{\label{T:O4}High-temperature UARs for $N=4$. The error bar for $r_8$ marked with an asterisk is under-estimated, see main text.}
	\begin{ruledtabular}
		\begin{tabular}{c  c c c|c c c}
			& \cite{toldin2003scaling}  & d=3 exp. & $\epsilon$ exp. & LPA & $\mathcal{O}(\partial^2)$ &$\mathcal{O}(\partial^4)$\\
			\hline 
			$g_4^+$   &    & 17.30(6)\cite{Guida:1998bx} &17.5(3)\cite{pelissetto2000effective}  & 20.0 & 17.2(7) & 17.31(4)\\
			
			$r_6$   &     &1.81(3)\cite{pelissetto2000effective,pelissetto1998effective}  &1.780(8)\cite{pelissetto2000effective}  & 1.73 & 1.80(2) & 1.809(3)\\
			
			$r_8$  &    &0.456\cite{sokolov1999renormalized}  & 0.2(4)\cite{pelissetto2000effective} & 0.485 & 0.51(1*) & 0.60(3) \\
			
			$r_{10}$  & -5(6)   &  &9(17)\cite{pelissetto2000effective}  & -5.7 & -10.0(15) & -9.6(4)\\
			
		\end{tabular}
	\end{ruledtabular}
\end{table}

\begin{table}
	\caption{\label{T:O5}High-temperature UARs for $N=5$. The error bar for $r_8$ marked with an asterisk is under-estimated, see main text.}
	\begin{ruledtabular}
		\begin{tabular}{c c|c c c}
			& \cite{butti2005critical}  & LPA & $\mathcal{O}(\partial^2)$ &$\mathcal{O}(\partial^4)$\\
			\hline
			$g_4^+$ & 15.74(2) & 17.9 & 15.8(5) & 15.77(3)\\
			& 15.6(1)  &&&\\
			$r_6$&  1.72(2)  & 1.65 & 1.73(2) & 1.739(2)\\
			& 1.70(1)  &&&  \\	
			$r_8$&  -1(3)   & 0.04 & 0.09(2*) & 0.16(2) \\
			& -0.3(5)  &&& \\
			$r_{10}$&  3(8) & -3.0 & -7.6(16) & -7.0(6)\\
			&   &&&\\
		\end{tabular}
	\end{ruledtabular}
\end{table}

The results shown in Tables \ref{T:O3}, \ref{T:O4} and \ref{T:O5} for the $O(3)$, $O(4)$ and $O(5)$ universality classes, respectively, complete our presentation of high-temperature UARs. The precision obtained for high temperature UARs using the DE at order $\mathcal{O}(\partial^4)$ is similar or better than the one obtained with many other methods. It is interesting to note that the precision achieved within the DE when $N$ grows tends to improve. This is expected because for all quantities that can be extracted from the potential $U$ and the $Z$ function, the DE becomes exact when $N\to \infty$ \cite{DAttanasio:1997yph}. A systematic study of the large-$N$ behavior of UARs goes beyond the scope of the present work and is left for a future analysis.

Before considering UARs that involve low temperature amplitudes, let us discuss the quality of the high-temperature UARs obtained. First, it is worth to mention that (putting aside the case of $r_8$ at order $\mathcal{O}(\partial^2)$ previously discussed) our estimates for high temperature UARs are systematically self-consistent in the sense that estimates at order $\mathcal{O}(\partial^4)$ are within error bars of order $\mathcal{O}(\partial^2)$. Second, in almost all cases the results are accurate in the sense that we are within error bars of most precise previous estimates. In some cases this is not strictly possible because for some UARs previous estimates in the literature are mutually incompatible. However, in all cases we are compatible with at least one of the most precise results in the literature. Moreover, our results are very precise. In many cases our error bars are of the same order of best previous estimates in the literature achieving, in many cases, the best precision at order $\mathcal{O}(\partial^4)$. In particular, our estimates for $r_6$, $r_8$ and $r_{10}$ tend to be the most precise in the literature. Finally, (again, putting aside the case of $r_8$) the precision of our results are systematically improved when going from order $\mathcal{O}(\partial^2)$ to order $\mathcal{O}(\partial^4)$.

\subsection{Other universal amplitude ratios}

So far we have described our results on the direct calculation of the high-temperature UARs $\{g_4,r_i\}$ through the numerical solution of the NPRG equations. We now turn to the remaining universal quantities discussed in Section 3. As we have explained in that section and in Appendix D, these quantities are calculated through the parametrization procedure. Our final results for these remaining UARs are presented in tables \ref{remUAR2}, \ref{remUAR3}, \ref{remUAR4} and \ref{remUAR5}. Let us now briefly review how these final results are obtained.\\
\indent As explained in more detail in Appendix D, we have implemented the parametrization method in five different ways, which we have labeled ($n=\{0,1A,1B,2A,2B\}$). These differ on the function which we have chosen to parametrize ($m(\theta)$ or $h(\theta)$) and on the order of the parametrization. The resulting UARs obtained for the different parametrization schemes  for $N=2$ and $N=3$ are presented in tables \ref{rawn2d2},\ref{rawn2d4},\ref{rawn3d2} and \ref{rawn3d4} of Appendix C. These tables also show the different results obtained from the parametrization procedure when taking the input data (critical exponents, $g_4$, $r_i$) from calculations at different order in the DE. The errors quoted in these tables correspond to those propagated through the parametrization method from the errors in the exponents, $g_4$ and the $r_i$.\\
\indent In order to extract our final results from this data, in principle we would like to use parametrizations $n=2A$ and $n=2B$, given that these involve the most $r_i$ coefficients (up to $r_{10}$) and thus carry the most information of the high-temperature phase. However, looking at the data in the aforementioned tables it is clear that the results of scheme $2A$ propagate a much higher error than those of scheme $2B$. This is likely due to the fact that scheme $2A$ has a higher dependence than $2B$ on $r_8$ and $r_{10}$, which are less precisely estimated in our method than $r_6$ and the critical exponents. This is in agreement with a similar observation done in \cite{campostrini2001critical}. For this reason we have chosen to present the result of scheme $2B$ as our final result. In order to provide a final error estimate, we take the RMS sum of the error propagated from the inputs to the parametrization scheme with the error associated to the parametrization procedure, which we estimate as the difference between the results of scheme $2B$ and scheme $1B$. The final results and error estimates are presented in tables \ref{remUAR2}, \ref{remUAR3}, \ref{remUAR4} and \ref{remUAR5} for the different orders in the DE.\\
\indent Finally, we would like to point out that while the error at $\mathcal{O}(\partial^4)$ in the derivative expansion is dominated by the error coming from the parametrization procedure, at $\mathcal{O}(\partial^2)$ this error is similar with that propagated from the error in the critical exponents and high-temperature UARs. This would appear to imply that a specialized study of the low-temperature phase done at $\mathcal{O}(\partial^2)$ 
would reduce error bars but would not have an effect as large as the one for the calculation of UARs at order $\mathcal{O}(\partial^4)$ in the DE. This may explain why, at in contrast with what happens for high temperature UARs, error bars of UARs involving low temperature amplitudes estimated at order $\mathcal{O}(\partial^4)$ are not substantially better than at order $\mathcal{O}(\partial^2)$. Indeed, the main source of error is not here the DE but the parametrization method and this explains the low apparent convergence of the DE for those quantities. Despite of this fact, estimates coming from the DE remain very precise even for UARs that require the use of the parametrization procedure, yielding in some cases the most precise results in the literature. Moreover, as for high temperature UARs, they remain accurate and self-consistent (in the sense that $\mathcal{O}(\partial^2)$ and $\mathcal{O}(\partial^4)$ are compatible within error bars).

Interestingly, while our results are in general compatible within error bars to most results in the literature, this does not appear to be the case for the UARs involving low-temperature quantities for $N=5$. Indeed, many of the results for these quantities from reference \cite{butti2005critical} are slightly incompatible with ours. This may be due to the fact that, even though the authors in \cite{butti2005critical} also use a parametrization procedure similar to the one employed in the present work, their central values for $r_8$ and $r_{10}$ are radically different from ours, and are compatible with ours only due to comparatively large error bars. Since these are inputs for the parametrization procedure, this could explain the difference found in the results. 

\begin{table*}
	\caption{\label{remUAR2}Other UARs for $N=2$.}
	\begin{ruledtabular}
		\begin{tabular}{ccccccc|ccc}
			& IHT--PR & HT & MC & $d$=3 exp & $\epsilon$ exp & experiments & LPA & $\mathcal{O}(\partial^2)$ & $\mathcal{O}(\partial^4)$ \\
			\hline
			
			$U_0$
			&1.062(4) \cite{campostrini2001critical} &  & 1.12(5) \cite{engels2000equation,cucchieri2002universal} & 
			1.056(4)\cite{larin1998five}&1.029(13)\cite{bervillier1986estimate} &1.053(2)\cite{He4exp} &1.48  &1.07(6)  &1.0596(74) \\
			&1.055(3) \cite{campostrini2000critical} &  & & 1.045 \cite{kleinert2001three}&   & 1.067(3) \cite{singasaas1984universality}&  &  & \\
			&        &  & & &  & 1.058(4) \cite{lipa1983very} &  &  &  \\
			&        & & &  &   & 1.088(7) \cite{takada1982critical} &  & &  \\

			$R_\alpha$
			&4.3(2) \cite{campostrini2001critical} &  & 4.20(5) \cite{cucchieri2002universal} & 4.39(26)
			\cite{strosser2000universal} & &4.154(22)\cite{He4exp}  &3.8  &4.1(2) & 4.0(2) \\
			&&&4.01(5)\cite{hasenbusch2006three}&&&&&&\\
			
			$R_\chi$ &  1.35(7) \cite{campostrini2001critical}  &    & 1.356(4) \cite{engels2000equation}    &  &  1.407   & &1.38  &1.4(1)  &1.36(7) \\
			
			$R_c$    
			&  0.127(6) \cite{campostrini2001critical}&                  &  0.128(2)\cite{hasenbusch2008monte}   &  0.123(3) \cite{strosser1999minimal}      &  0.106  & & 0.192 &0.13(2) &0.128(7) \\

			$R_4$ &  7.5(2) \cite{campostrini2001critical}  &                 &     &  &  & & 8.0 &7.5(4) &7.5(2) \\
			
			$R_\xi^+$ 
			& 0.355(3) \cite{campostrini2001critical} & 0.361(4) \cite{butera1999critical}& 0.3562(10)\cite{hasenbusch2008monte} &
			0.3606(20)\cite{bagnuls1985nonasymptotic,bervillier1980universal}& 0.36 \cite{bervillier1976universal}  & &0.392  &0.35(1)  &0.356(3) \\
			
			&    &  0.130 \cite{kleinert2001three}      &  & &  &  &\\
			
			$P_m$&&&&&&&1.19(1)&1.19(2)&1.190(16)\\
			
			$R_p$&&&&&&&2.03&1.990(6)&1.987(1)\\
		\end{tabular}
	\end{ruledtabular}
\end{table*}

\begin{table*}
	\caption{\label{remUAR3}Other UARs for $N=3$.}
	\begin{ruledtabular}
		\begin{tabular}{cccccc|ccc}
			& IHT--PR \cite{Campostrini_2002} & $d$=3 exp & $\epsilon$ exp & HT & experiments&LPA&$\mathcal{O}(\partial^2)$& $\mathcal{O}(\partial^4)$\\
			\hline
			$U_0$ & 1.56(4) & 1.51(4) \cite{larin1998five} & 1.521(22) \cite{bervillier1986estimate}  
			& & 1.50(5) \cite{kaul1994asymptotic}&1.90&1.52(6)&1.50(2)\\
			&         & 1.544 \cite{kleinert2001three} & & &    1.27(9) \cite{marinelli1996effect} &1.823 \cite{Berges:1995mw,Berges:2000ew}&&\\
			& & & &  & 1.4(4) \cite{ramos2001spontaneous}&&& \\ 
			
			$R_\alpha$ & 4.3(3)  &  4.4(4) \cite{larin1998five,Campostrini_2002}  & 4.56(9)
			\cite{bervillier1986estimate}  && &3.2&3.8(3)&3.8(2) \\
			&         &  4.46 \cite{kleinert2001three,Campostrini_2002}  &   &&  &3.41 \cite{Berges:1995mw,Berges:2000ew}&&\\
			
			$R_\chi$ &  1.31(7) &  & 1.33 \cite{abe1978note}  & & &1.17&1.20(9)&1.20(5)\\
			&&&&&&1.11 \cite{Berges:1995mw,Berges:2000ew}&&\\
			
			$R_C$    & 0.185(10) & 0.189(9) \cite{strosser1999minimal} & 0.17 \cite{aharony1976universal}  & & &0.298&0.20(2)&0.202(7)\\
			& & 0.194 \cite{kleinert2001three} & &  & &&&\\

			$R_4$ & 7.8(3) & &&&  &7.97&7.4(3)&7.4(2)\\
			
			$R_\xi^+$ & 0.424(3) & 0.4347(20) \cite{bagnuls1985nonasymptotic} & 0.42 \cite{bervillier1976universal} &
			0.431(5) \cite{butera1999critical} & &0.48&0.43(1)&0.427(2)\\
			& & 0.4319(17) \cite{bervillier1980universal} &  & 0.433(5) \cite{butera1999critical}& &&&\\ 
			
			$P_m$ & 1.18(2) & &&&  &1.13&1.15(2)&1.15(1)\\
			
			
			$R_p$ & 2.020(6) & && & &2.06&2.013(7)&2.012(2)\\
		\end{tabular}
	\end{ruledtabular}
\end{table*}

\begin{table}\centering
	\caption{\label{remUAR4}Other UARs for $N=4$.}	
	\begin{ruledtabular}
		\begin{tabular}{c c c c |c c c}
			& \cite{toldin2003scaling} &\cite{cucchieri2005equation} & MC\cite{engels2000goldstone,engels2001finite}  & LPA & $\mathcal{O}(\partial^2)$ &$\mathcal{O}(\partial^4)$\\
			\hline 
			
			$U_0$  & 1.91(10)&1.8(2) &   &2.02  &1.84(5)  &1.82(4) \\
			
			$R_{\chi}$  & 1.12(11)&1.10(5) &1.126(9)    &1.04  &1.08(7)  &1.087(32)\\
			
			$R_c$  &0.27(2)& 0.26(1) &   &0.38  &0.27(2)  &0.273(7)\\
			
			$R_4$  & 7.6(4)& &8.6(9)   &8.00  &7.4(3)  &7.4(1)\\
			
			$R_{\alpha}$  & & &    &2.4  &3.4(3)  &3.4(2)\\
			
			$R_{\xi}^+$  &  &&    &0.53  &0.49(1)  &0.488(2)\\
			
			$P_m$  &1.13(2)&  &1.11(2)   &1.11  &1.12(2)  &1.123(9)\\
			
			$R_p$  & 2.042(7)& &   &2.08  &2.038(7)  &2.037(2)\\
		\end{tabular}
	\end{ruledtabular}
\end{table}

\begin{table}
	\caption{\label{remUAR5}Other UARs for $N=5$.}
	\begin{ruledtabular}
		\begin{tabular}{c c |c c c}
			& \cite{butti2005critical} & LPA & $\mathcal{O}(\partial^2)$ &$\mathcal{O}(\partial^4)$\\
			\hline
			$U_0$& 2.2(2) &1.99  &2.04(7)  &1.86(9)\\
			
			$R_{\chi}$& 1.2(1)  &0.98  &1.01(5)  &0.96(4)\\
			
			$R_c$& 0.28(2) &0.44  &0.34(2)  &0.341(8)\\
			
			$R_4$& 8.3(5)  &8.2  &7.5(2)  &7.1(2)\\
			
			$R_{\alpha}$& 4.2(6) &1.93  &2.9(3)  &2.5(3)\\
			
			$R_{\xi}^+$& 0.527(6)  &0.58  &0.54(1)  &0.519(1)\\
			
			$P_m$& 1.15(3)  &1.09  &1.10(1)  &1.09(1)\\
			
			$R_p$& 2.069(9)  & 2.11 &2.062(7)  &2.039(2)\\
		\end{tabular}
	\end{ruledtabular}
\end{table}

\section{Conclusions}

In this work we have used the DE of the NPRG to order $\mathcal{O}(\partial^4)$ to calculate the Universal Amplitude Ratios for three-dimensional Ising and $O(N)$ universality classes.
Recently developed techniques, based on the existence of a small parameter in the DE, allow for the estimation of error bars \cite{Balog:2019rrg,DePolsi:2020pjk}. The resulting precision is competitive with best estimates for UARs in the literature \cite{Pelissetto02}, in some cases obtaining the most precise estimate to date. The results, moreover, turn out to be accurate in the sense that in almost all cases they are compatible with most precise results in the literature, whenever these are more precise than ours.

The calculation of UARs is interesting, first, because of their empirical interest. This has led to their study by various theoretical methods and, in some cases, also to their experimental determination \cite{Pelissetto02}.  At the same time, the study of UARs is challenging because it requires knowledge of the theory beyond the linear regime around the Renormalization Group fixed point.  These quantities are universal but require the study of RG trajectories that go towards the high or low temperature phases. The ability to calculate these types of properties shows to be a strength of the NPRG that proves to be not only an accurate method but also an all-terrain one.

In the present work we have avoided studying the trajectories that lead to the low temperature phase using the ``parameterization method'', which had already been established in the literature \cite{Pelissetto02}. This allows for the calculation of low temperature amplitudes from the knowledge of amplitudes in the high temperature phase (together with critical exponents). For UARs that involve low temperature amplitudes, the parametrization method introduces uncertainties that, roughly speaking, are of the same order than DE at order $\mathcal{O}(\partial^2)$, so at that order it is probably unnecessary to study the (numberically difficult) low temperature phase. However, it is interesting to note that
this procedure turns out to be our main source of error at order $\mathcal{O}(\partial^4)$ (the parametrization method ends up being less precise than DE at that order). 
To avoid this difficulty it is possible to use the DE for the study of the low temperature phase directly without using the parametrization method. However, this is much more demanding from the numerical point of view. Given the results obtained in the present work, this study becomes necessary and we plan to carry it out in the near future. This particularly important for the calculation of the UAR denoted by $U_0$ for $N=2$ where a particularly precise experimental determination is, for the moment, more precise than all theoretical estimates \cite{lipa1996heat,lipa2000specific}.\footnote{A similar statement can be made for $R_{\alpha}$ since this quantity is built out of $U_0$ and the critical exponent $\alpha$.}

The present study also opens up other new perspectives. In particular, we have restricted ourselves in this work to the three-dimensional case, but it is natural to extend the analysis to the two-dimensional case. Likewise, in the particular case of the Ising universality class, the DE has been extended to order $\mathcal{O}(\partial^6)$ for the calculation of critical exponents \cite{Balog:2019rrg} and it is natural to apply this analysis also for the calculation of UARs. Concerning, the Ising universality class, it is worth mentioning again that there exists in the literature a variety of different implementations of the parametrization method, differing in what type of functions are used and how the parameters are determined (see for instance \cite{Pelissetto02} for a discussion), which warrants a more detailed examination of this technique. As a consequence, we have left for future work an analysis of UARs for this universality class concerning low temperature amplitudes.

There are also perspectives concerning other values of $N$. We did not perform a study of the the large-$N$ limit in the present work, as the one performed in Ref.~\cite{DePolsi:2020pjk} for critical exponents but a systematic analysis of the large-$N$ behavior of the UARs within DE should be performed in the future. As recalled in Section~\ref{results}, in the large-$N$ limit the DE becomes exact for quantities that can be extracted from the potential $U$ and the $Z$ function. This includes all UARs studied in the present article. An analysis of the large $N$ behavior (including $1/N$ corrections \cite{butti2005critical}) would be interesting. Moreover, we did not analyse the case of $N\to 0$ where other kind of UARs are of interest and could be calculated with the present techniques \cite{Pelissetto02}.

Finally, from the point of view of the methodology employed, this work shows the capacity of the DE to obtain precision results for quantities that are not exclusively dominated by the fixed point of the RG. This is in a sense just the tip of the iceberg because the same methodology can be used to calculate other large-distance quantities that do not even have to be universal. For example, in Refs.~\cite{Machado:2010wi,CAILLOL2012291}, the NPRG and approximations close to the DE were successfully employed for the calculation of critical temperatures. Extending the use of the DE (or similar approximations) at high order for the calculation of such properties seems a challenging but not unapproachable goal.

\begin{acknowledgments}
	
	The authors thank I. Balog, B. Delamotte, A. Ran\c{c}on and M. Tissier, for a very careful reading of a previous version of this manuscript.
	This work was supported by Grant I+D number 412 of the CSIC (UdelaR) Commission and Programa de Desarrollo de las Ciencias B\'asicas (PEDECIBA). GH would like to thank specifically the PEDECIBA
	program for funding his postdoc scholarship. 
	
\end{acknowledgments}

\appendix
\section{Numerical procedure} \label{App:Num}

In this appendix we give the details of the numerical procedure in order to compute the different UARs. We start by describing the steps for obtaining the flow equations for the effective action $\Gamma_k$. We proceed to present the treatment of these equations and specify the numerical parameters to obtain the fixed point solution. Finally we give the details for the RG time integration procedure which allows us to flow away from the fixed point and compute the different UARs. 

\subsection{Derivation of the flow equations}

When computing thermodynamic quantities using a given order of the DE of the NPRG, we need to obtain the equations describing the evolution of the functions in the ansatz as we integrate the fluctuations at different scales. To do so, we proceed as follows. We differentiate $n$ times the NPRG equation Eq.~\eqref{wettericheq}. The resulting expression will depend on the $\Gamma_k^{(n+1)}$ and $\Gamma_k^{(n+2)}$ vertex and, therefore, we insert the ansatz at a given order of the DE and compute the $l$-point vertex $\Gamma_k^{(l)}$ with $l$ up to $n+2$. We evaluate in a uniform field configuration and take its Fourier transform. We then insert these functions into the differentiated Eq.~\eqref{wettericheq} which results in many terms which are proportional to different momenta and colour indices structures. 

We then must identify the coefficient of these structures between left- and right-hand sides to obtain the evolution of the different functions of the ansatz. However, before doing so, if we are implementing a given order $\mathcal{O}(\partial^b)$ of the DE, we truncate the expression obtained so far (from the vertex functions) in the right-hand side at order $b$ in the internal and external momenta \textit{before} expanding the propagators. We emphasize this distinction because up to recent years it was common to keep all terms after plugging in the ansatz into the vertex functions. The difference between these truncations corresponds to higher order corrections in the DE.

\subsection{Finding the fixed point solution}

Once the equations are obtained, we seek for its fixed point solution. We do this by finding the zeros of the beta functions starting from a reasonable close solution. This method was possible due to previous works from where we already have reasonable fixed point solutions to start with. We refer to Ref.~\cite{DePolsi:2020pjk} for a description on how to obtain these approximate solutions. Once we solve for a certain set of parameters, say the regulator scale $\alpha$ or the dimension of the order parameter $N$, we lean on the fact that the equations behaves smoothly on these parameters. This implies that we can use the solution we found for a set of parameters in order to find a solution with a different, but close, set of parameters. To find the fixed point solution we implemented a Newton-Raphson algorithm which serves just well enough for our purposes.

In all the calculations, we used a discretization of $\rho$ into a grid with $N_\rho=40$ points. With a $\rho$ step which kept the minima of the potential around a fourth part of the $\rho$ box size. As a consequence, different sizes of the $\rho$ box were used for different values of $N$. Additionaly, we choose as the renormalization condition $\tilde{Z}(\tilde{\rho_{i}})|_{i=N_{\rho}/4}=1$, being $\tilde{Z}(\tilde{\rho})$ the dimensionaless version of the ansatz function $Z_k(\rho)$. For the momentum integrals we use the method provided in the quadpack library corresponding to an adaptative 21 point Gauss-Kronrod quadrature rule (qags). The derivatives with respect to $\rho$ were approximated using $7$-points centered discretization which made more stable the computation of quantities such as $r_{10}$ with respect to a $5$-point rule. The derivatives at the boundary of the $\rho$ box were also computed with $7$-points but not centered for obvious reasons.

\subsection{Linear stability analysis of the fixed point}

In order to compute critical exponents and the eigendirections around the fixed point we performed a linear stability analysis of the fixed point solution. To do this, we start from the very precise fixed point solution after performing the Newton-Raphson method and compute the stability matrix $\mathcal{M}$ by evaluating the beta functions at perturbed points from this solution. From the stability matrix $\mathcal{M}$ we perform a standard linear stability analysis in order to obtain the critical exponents (related to the eigenvalues) and the eigendirections. The later is important for the next step as will be clear shortly.

\subsection{Flowing to the high-temperature regime}

From the fixed point solution and the eigendirections we perturbate the system into the high-temperature regime by taking a perturbation along the unstable eigendirection which corresponds to the leading eigenvalue. We make sure to pertubate into the high-, and not low-, temperature regime and we add to this perturbation, another one in the first correction or least stable eigendirection. This is done in order to approach the fixed point along the same eigendirection that a typical microscopic system would under the NPRG. One does retain, in this way, as much physics as possible. This is useful in general because if one perturbates only along the relevant eigendirection the correction to scaling would be suppress and, for example, the critical exponent $\omega$ could not be computed from the flow. We made sure to verify the convergence on the strength of the perturbation. This perturbation position us along a path which coincides with the path that a general system would follow if one is to integrate the flow equations from a microscopic theory. For the time integration of the flow equations we used a fourth order Runge-Kutta algorithm using a fixed RG time step $ds=-1\times10^{-3}$ and with free boundary contidion for the $\rho$ box.

\section{Expressions for high-temperature UARs}\label{appendixhifhtUar}

The equation of state at high temperature is written in terms of the potential $U(\rho)$ appearing in \eqref{ansatz-order4} evaluated at constant magnetization

\begin{equation}
H_i = \partial_{M^i}U = M_i \partial_{\rho}U 
\end{equation}

\noindent so that 

\begin{equation}
H = M \partial_{\rho}U.
\label{EOSinU}
\end{equation}

Let us now take 

\begin{equation}
U = \displaystyle\sum_{n=0}U^{(n)}\frac{\rho^n}{n!}.
\end{equation}

In these terms it is straightforward to write

\begin{align}
H &= \frac{(U^{(1)})^{3/2}\sqrt{2}}{3!^{1/2}(U^{(2)})^{1/2}}\left(z +\frac{z^3}{3!} \right.\nonumber\\
&+\left.\displaystyle\sum_{n=3}\frac{(2n-1)!}{(n-1)!3!^{n-1}}\frac{(U^{(1)})^{n-2}U^{(n)}}{(U^{(2)})^{n-1}}\frac{z^{2n-1}}{(2n-1)!}  \right)
\label{GfromU}
\end{align}

\noindent where 

\begin{equation}
z = \sqrt{\frac{U^{(2)}3!}{U^{(1)}2}}M
\end{equation}

\noindent Notice that equation \eqref{GfromU} is precisely of the form of \eqref{highTEOS} and indeed the term in parenthesis in \eqref{GfromU} coincides with the function $G(z)$ defind in \eqref{highTEOS}. From this we immediately find that 

\begin{align}\label{r6eq}
r_6 &= \frac{5}{3}\frac{U^{(3)}U^{(1)}}{(U^{(2)})^{2}}\\ \label{r8eq}
r_8 &= \frac{35}{9}\frac{U^{(4)}(U^{(1)})^2}{(U^{(2)})^{3}}\\ \label{r10eq}
r_{10} &= \frac{35}{3}\frac{U^{(5)}(U^{(1)})^3}{(U^{(2)})^{4}}
\end{align}

\noindent Notice that in contrast with what happens with $g^+_4$, the $r_i$ do not depend on $Z$ but rather can be calculated exclusively from $U$.

\section{The parametrization procedure}\label{appParam}

The strategy we will employ starts by approximating the functions $m(\theta)$ and $h(\theta)$ by polynomials. Since by redefinition of $\theta$ one of these two functions can always be taken to be a fixed known function, this defines two possible approximation schemes:\\
\indent Scheme A:

\begin{align}
m(\theta) &= \theta\left(1 + \displaystyle\sum_i^n c_i\theta^{2i}\right)\\
h(\theta) &= \theta \left(1-\theta^2/\theta_0^2\right)^2
\end{align}

Scheme B:

\begin{align}
m(\theta) &=\theta\\
h(\theta) &=\theta \left(1-\theta^2/\theta_0^2\right)^2\left(1 + \displaystyle\sum_i^n c_i\theta^{2i}\right)
\end{align}

\noindent Notice that we have already imposed that $h(\theta)$ has a zero at $\theta_0$, and indeed a double zero as must be the case for $N>1$ systems in $d=3$ (see for instance \cite{Pelissetto02} for discussion on this last point). There are then $n+2$ free parameters to be determined in both schemes: the $c_i$, $\theta_0$ and $\sigma$ from \eqref{Goftheta}. As mentioned in the main text, there exist in the literature various strategies to accomplish this. Following \cite{Campostrini_2002} we will impose that our parametrization reproduces the coefficients in the low-$z$ polynomial expansion of $G(z)$. It is straightforward to see that this is automatically satisfied in the parametrization chosen at order $\mathcal{O}(z)$. Thus to have $n+2$ equations to be used to fix $n+2$ quantities, we will require that our parametrization reproduces the coeffcients of $G(z)$ up to order $\mathcal{O}(z^{2n+5})$. This restricts the number o $r_i$ that are used to fix the coefficients of our ansatz at a given order $n$ as shown in the following table. 
\begin{center}
	\begin{tabular}{|c|c|}
		\hline
		$n=0$& $r_6$ \\
		\hline
		$n=1$& $r_6$, $r_8$ \\
		\hline
		$n=2$&  $r_6$, $r_8$, $r_{10}$\\
		\hline
	\end{tabular} 
\end{center}
In practice, we impose these conditions by expanding equation \eqref{Goftheta} in powers of $\theta$ to order $\mathcal{O}(\theta^{2n+5})$. This yields an algebraic system of equations which is easily solved (numerically) for the $n+2$ coefficients defining our parametrization. Doing this for $n=0,1,2$ for both of the schemes A and B yields five different schemes which we label $nA$ and $nB$ (notice that schemes $0A$ and $0B$ are equivalent).\\
\indent Once the $n+2$ coefficients are determined the approximate representation for the EOS is obtained and the remaining UARs easily follow. In fact, there exist in the literature formulae to obtain the remaining UARs directly from he functions defining the parametrization (see for example \cite{Campostrini_2002}). We reproduce the relevant ones here for completeness.\\
\indent Let us define first two auxiliary functions $Y(\theta)$ and $g(\theta)$. The first is given by

\begin{equation}
Y(\theta) = (1-\theta^2)m'(\theta) + 2\beta \theta m(\theta)
\end{equation}

\noindent whereas $g$ is obtained as the solution of the differential equation

\begin{equation}
(1-\theta^2)g'(\theta) + 2(2-\alpha)\theta g(\theta) = Y(\theta) h(\theta)
\end{equation}

\noindent fixed by imposing regularity at $\theta = 1$. Furthermore we will require the value $\theta_M$ defined by

\begin{equation}
\beta \delta F(z(\theta_M))F''(z(\theta_M))-\gamma F'(z(\theta_M))^2=0
\end{equation}

In terms of these the UARs are obtained through the following formulas

\begin{align}
U_0 &= (\theta_0^2-1)^{2-\alpha}\frac{g(0)}{g(\theta_0)}\\
R_{\chi} &= (\theta_0^2-1)^{-\gamma}m(\theta_0)^{\delta -2}m(1)^{-\delta}h(1)\\
R_C &= -\alpha (1-\alpha)(2-\alpha)(\theta_0^2-1)^{2\beta}m(\theta_0)^{-2}g(0)\\
R_4 &= \sigma^2 m(\theta_0)^2 (\theta_0^2-1)^{-2\beta}
\end{align}

\noindent where $f$ and $G$ are obtained through the use of \eqref{foftheta} and \eqref{Goftheta} and $z_M$ and $x_M$ are obtained from $\theta_M$ through the use of \eqref{xoftheta} and \eqref{zoftheta}.

\section{Raw data}\label{Ap:NumParam}

\begin{table}[h!]
	\caption{\label{table_supplN1}  Raw data for $N=1$ high-temperature UARs in $d=3$ obtained with the considered regulators at various orders of the DE. Quantities marked with a $*$ are computed with an extension of the PMS criterion.}
	\begin{ruledtabular}
		\begin{tabular}{llllll}
			& \hspace{-0.5cm}regulator          &  $g_4$         &  $r_6$ &  $r_8$ &  $r_{10}$ \\
			\hline
			LPA   &$W$                                &    29.15      &   2.01     &  2.65    &  -9.5   \\
			&$E$                                &     29.23     &   2.00     &  2.63    &   -9.5		\\          
			\hline
			$O(\partial^2)$
			&$W$                                 &    23.09      & 2.05  &  2.40    & -14.8  \\
			&$E$                                 &    23.05      & 2.05  &   2.40   & -14.8 \\
			\hline
			$O(\partial^4)$
			&$W$                                &    23.590      & 2.063  &  2.586  & -14.11 \\
			&$E$                                &    23.604      & 2.065  &  2.604   & -14.05  \\
		\end{tabular}
	\end{ruledtabular}
\end{table}
\begin{table}[h!]
	\caption{\label{table_supplN2}  Raw data for $N=2$ high-temperature UARs in $d=3$ obtained with the considered regulators at various orders of the DE. Quantities marked with a $*$ are computed with an extension of the PMS criterion.}
	\begin{ruledtabular}
		\begin{tabular}{llllll}
			& \hspace{-0.5cm}regulator          &  $g_4$         &  $r_6$ &  $r_8$ &  $r_{10}$ \\
			\hline
			LPA   &$W$                                &     25.64     &    1.91    &   1.79   &   -9.5  \\
			&$E$                                &    25.70      &    1.91    &   1.78   &   -9.5 	\\          
			\hline
			$O(\partial^2)$
			&$W$                                 &     20.82     & 1.96  &   1.64   &  -14.1 \\
			&$E$                                 &     20.79     & 1.96  &   1.64   &  -14.2 \\
			\hline
			$O(\partial^4)$
			&$W$                                &    21.174      & 1.971  &  1.795  & -13.55 \\
			&$E$                                &     21.177     & 1.972  &  1.810   &  -13.52 \\
		\end{tabular}
	\end{ruledtabular}
\end{table}
\begin{table}[h!]
	\caption{\label{table_supplN3}  Raw data for $N=3$ high-temperature UARs in $d=3$ obtained with the considered regulators at various orders of the DE. Quantities marked with a $*$ are computed with an extension of the PMS criterion.}
	\begin{ruledtabular}
		\begin{tabular}{llllll}
			& \hspace{-0.5cm}regulator          &  $g_4$         &  $r_6$ &  $r_8$ &  $r_{10}$ \\
			\hline
			LPA   &$W$                                &     22.59     &    1.81    &    1.07  &  -8.1   \\
			&$E$                                &    22.63      &   1.81     &  1.06    &  -8.0  	\\          
			\hline
			$O(\partial^2)$
			&$W$                                 &    18.90      & 1.88  &   1.02   & -12.5  \\
			&$E$                                 &    18.88      & 1.88  &   1.02   & -12.5 \\
			\hline
			$O(\partial^4)$
			&$W$                                &    19.103      & 1.886*  &  1.136  & -12.15 \\
			&$E$                                &    19.092      & 1.886*  &  1.144   & -12.14  \\
		\end{tabular}
	\end{ruledtabular}
\end{table}
\begin{table}[h!]
	\caption{\label{table_supplN4}  Raw data for $N=4$ high-temperature UARs in $d=3$ obtained with the considered regulators at various orders of the DE. Quantities marked with a $*$ are computed with an extension of the PMS criterion.}
	\begin{ruledtabular}
		\begin{tabular}{llllll}
			& \hspace{-0.5cm}regulator          &  $g_4$         &  $r_6$ &  $r_8$ &  $r_{10}$ \\
			\hline
			LPA   &$W$                                &    20.02      &  1.73      &  0.49    &  -5.7   \\
			&$E$                                &    20.05      &    1.72    &  0.48    &    -5.6		\\          
			\hline
			$O(\partial^2)$
			&$W$                                 &    17.23*      & 1.80*  &   0.50*   & -10.0  \\
			&$E$                                 &     17.22*     & 1.80*  &   0.51*   & -10.1 \\
			\hline
			$O(\partial^4)$
			&$W$                                &     17.319     & 1.808*  &  0.593*  &  -9.61 \\
			&$E$                                &      17.298    & 1.809*  &  0.597*   &  -9.65 \\
		\end{tabular}
	\end{ruledtabular}
\end{table}
\begin{table}[h!]
	\caption{\label{table_supplN5}  Raw data for $N=5$ high-temperature UARs in $d=3$ obtained with the considered regulators at various orders of the DE. Quantities marked with a $*$ are computed with an extension of the PMS criterion.}
	\begin{ruledtabular}
		\begin{tabular}{llllll}
			& \hspace{-0.5cm}regulator          &  $g_4$         &  $r_6$ &  $r_8$ &  $r_{10}$ \\
			\hline
			LPA   &$W$                                &    17.88      &   1.65     &   0.05   &   -3.0  \\
			&$E$                                &     17.90     &    1.65    &  0.03    &    -3.0		\\          
			\hline
			$O(\partial^2)$
			&$W$                                 &    15.77      & 1.73*  &  0.085*    &  -7.5 \\
			&$E$                                 &    15.77      & 1.74*  &  0.089*    & -7.7 \\
			\hline
			$O(\partial^4)$
			&$W$                                &     15.781     & 1.738*  &  0.155*  & -6.90 \\
			&$E$                                &     15.754     & 1.739*  &  0.159*   &  -7.00 \\
		\end{tabular}
	\end{ruledtabular}
\end{table}

\begin{table*}
	
	\begin{tabular}{|c|c|c|c|c|c|}
		\hline
		& \text{n=0} & \text{n=1A} & \text{n=1B} & \text{n=2A} & \text{n=2B} \\
		\hline
		$U_0$ & 1.084(74) & 1.068(58) & 1.074(64) & 1.074(63) & 1.072(62) \\\hline
		$R_{\alpha }$ & 4.819(66) & 3.90(20) & 4.23(11) & 4.25(47) & 4.14(21) \\\hline
		$R_{\chi }$ & 1.637(61) & 1.15(12) & 1.430(65) & 1.94(95) & 1.39(11) \\\hline
		$R_C$ & 0.1037(95) & 0.147(19) & 0.121(11) & 00.99(39) & 0.125(14) \\\hline
		$R_4$ & 8.36(14) & 6.80(41) & 7.67(19) & 9.0(2.3) & 7.54(33) \\\hline
		$R_{\xi }$ & 0.346(12) & 0.364(12) & 0.355(11) & 0.350(19) & 0.356(12) \\\hline
		$P_M$ & 1.251(10) & 1.137(31) & 1.205(13) & 1.30(16) & 1.195(23) \\\hline
		$P_C$ & 0.3766(29) & 0.3766(29) & 0.3766(29) & 0.3766(29) & 0.3766(29) \\\hline
		$R_P$ & 1.9899(57) & 1.9899(57) & 1.9899(57) & 1.9899(57) & 1.9899(57)\\
		\hline
	\end{tabular}
	\caption{UARs for $N=2$ at $O(\partial^2)$ for the different parametrizations.}
	\label{rawn2d2}
\end{table*}

\begin{table*}
	\begin{tabular}{|c|c|c|c|c|c|}
		\hline
		& \text{n=0} & \text{n=1A} & \text{n=1B} & \text{n=2A} & \text{n=2B} \\\hline
		$U_0$ & 1.0708(85) & 1.0562(67) & 1.0618(74) & 1.0590(71) & 1.0596(71) \\\hline
		$R_{\alpha }$ & 4.789(16) & 3.803(75) & 4.177(32) & 3.98(13) & 4.028(63) \\\hline
		$R_{\chi }$ & 1.654(10) & 1.112(48) & 1.433(14) & 1.52(20) & 1.369(23) \\\hline
		$R_C$ & 0.1026(11) & 0.1522(61) & 0.1214(16) & 0.120(13) & 0.1279(26) \\\hline
		$R_4$ & 8.397(32) & 6.65(17) & 7.658(48) & 7.89(57) & 7.453(81) \\\hline
		$R_{\xi }$ & 0.3439(12) & 0.3629(21) & 0.3528(12) & 0.3554(40) & 0.3557(15) \\\hline
		$P_M$ & 1.2541(16) & 1.126(13) & 1.2053(27) & 1.225(43) & 1.1904(51) \\\hline
		$P_C$ & 0.3787(10) & 0.3787(10) & 0.3787(10) & 0.3787(10) & 0.3787(10) \\\hline
		$R_P$ & 1.9872(18) & 1.9872(18) & 1.9872(18) & 1.9872(18) & 1.9872(18)\\\hline
	\end{tabular}
	
	\caption{UARs for $N=2$ at $O(\partial^4)$ for the different parametrizations.}
	\label{rawn2d4}
\end{table*}

\begin{table*}
	\caption{\label{rawn3d2}UARs for $N=3$ at $O(\partial^2)$ for the different parametrizations.}
	\begin{ruledtabular}
		\begin{tabular}{c|ccccc}
			& \text{n=0} & \text{n=1A} & \text{n=1B} & \text{n=2A} & \text{n=2B} \\\hline
			$U_0$ & 1.622(82) & 1.489(49) & 1.539(63) & 1.528(63) & 1.523(56) \\\hline
			$R_{\alpha }$ & 4.52(13) & 3.56(27) & 3.92(17) & 3.84(49) & 3.80(26) \\\hline
			$R_{\chi }$ & 1.373(55) & 1.035(99) & 1.232(57) & 1.43(50) & 1.198(710) \\\hline
			$R_C$ & 0.177(15) & 0.227(26) & 0.196(16) & 0.178(52) & 0.201(19) \\\hline
			$R_4$ & 8.07(15) & 6.80(37) & 7.50(18) & 8.1(1.6) & 7.37(28) \\\hline
			$R_{\xi }$ & 0.423(12) & 0.434(13) & 0.427(12) & 0.424(17) & 0.428(12) \\\hline
			$P_M$ & 1.196(11) & 1.107(27) & 1.160(13) & 1.20(11) & 1.151(19) \\\hline
			$P_C$ & 0.3615(34) & 0.3615(34) & 0.3615(34) & 0.3615(34) & 0.3615(34) \\\hline
			$R_P$ & 2.0133(65) & 2.0133(65) & 2.0133(65) & 2.0133(65) & 2.0133(65)\\
		\end{tabular}
	\end{ruledtabular}
\end{table*}

\begin{table*}
	\caption{\label{rawn3d4}UARs for $N=3$ at $O(\partial^4)$ for the different parametrizations.}
	\begin{ruledtabular}
		\begin{tabular}{c|ccccc}
			& \text{n=0} & \text{n=1A} & \text{n=1B} & \text{n=2A} & \text{n=2B} \\\hline
			$U_0$ & 1.606(10) & 1.4724(94) & 1.5235(85) & 1.499(15) & 1.5039(93) \\\hline
			$R_{\alpha }$ & 4.518(17) & 3.520(62) & 3.901(29) & 3.71(11) & 3.755(54) \\\hline
			$R_{\chi }$ & 1.3926(88) & 1.029(27) & 1.243(10) & 1.31(11) & 1.201(15) \\\hline
			$R_C$ & 0.1751(19) & 0.2289(54) & 0.1949(21) & 0.190(12) & 0.2015(29) \\\hline
			$R_4$ & 8.125(27) & 6.77(11) & 7.533(39) & 7.74(35) & 7.376(61) \\\hline
			$R_{\xi }$ & 0.4207(14) & 0.4329(16) & 0.4252(13) & 0.4258(210) & 0.4269(14) \\\hline
			$P_M$ & 1.1999(16) & 1.1051(75) & 1.1625(22) & 1.180(25) & 1.1516(37) \\\hline
			$P_C$ & 0.36246(83) & 0.36246(83) & 0.36246(83) & 0.36246(83) & 0.36246(83) \\\hline
			$R_P$ & 2.0123(16) & 2.0123(16) & 2.0123(16) & 2.0123(16) & 2.0123(16)\\
		\end{tabular}
	\end{ruledtabular}
\end{table*}

\bibliographystyle{apsrev4-1} 
\bibliography{UAR_article}

\end{document}